\documentclass[10pt,letterpaper]{article}
\usepackage{setspace}
\usepackage{amsfonts}
\usepackage{booktabs}
\usepackage{siunitx}
\usepackage{multirow}
\usepackage{graphicx}
\usepackage{listings}
\usepackage{commath}
\usepackage{epstopdf}
\usepackage{xcolor}
\usepackage{url}
\usepackage{mathtools}
\usepackage{caption}
\usepackage{alltt}
\usepackage{mathrsfs}
\usepackage{float}
\usepackage{subcaption}
\usepackage{graphicx,fancyvrb}
\usepackage[colorinlistoftodos]{todonotes}
\usepackage[linesnumbered,ruled,vlined]{algorithm2e}
\usepackage{algpseudocode}
\usepackage{amsmath,amssymb}
\usepackage{booktabs}
\usepackage{caption}
\usepackage{float}
\usepackage{capt-of}
\usepackage{xcolor}

\usepackage{array}
\usepackage{arydshln}
\usepackage{todonotes}

\definecolor{mygreen}{rgb}{0,0.6,0}
\definecolor{mygray}{rgb}{0.5,0.5,0.5}
\definecolor{mymauve}{rgb}{0.58,0,0.82}

\usepackage{listings}
\lstnewenvironment{VerbatimText}[1][]{
    
    \lstset{fancyvrb=true,basicstyle=\footnotesize,captionpos=b,xleftmargin=2em,#1}
}{}

\newcommand{\mypm}{\mathbin{\tikz [x=1.4ex,y=1.4ex,line width=.1ex] \draw (0.0,0) -- (1.0,0) (0.5,0.08) -- (0.5,0.92) (0.0,0.5) -- (1.0,0.5);}}%
\newcommand{\fge}{{\bf {FGE}}}

\newcommand{\cmit}{{\bf {MIT}}}
\newcommand{\icmit}{{\bf {FNMIT}}}

\newcommand{\hcmit}{{\bf {HyMIT}}}

\newcommand{\pdal}{{\bf {CD}}}

\newcommand{\ate}{{\tt \textsc{ATE}}}
\newcommand{\nde}{{\tt \textsc{NDE}}}

\newcommand{\E}{{\tt \mathbb{E}}}
\newcommand{\pr}{{\tt \mathrm{Pr}}}
\newcommand{\sys}{{\textsc{HypDB}}}

\newcommand{\bigCI}{\mathrel{\text{\scalebox{1.07}{$\perp\mkern-10mu\perp$}}}}

\newcommand{\mc}[1]{\mathcal{#1}}

\newcommand{\ignore}[1]{}

\newcommand{\frv}[1]{{{{#1}}}}
\newcommand{\srv}[1]{{{{#1}}}}
\newcommand{\trv}[1]{{{{#1}}}}

\newcommand{\dan}[1]{{{\color{magenta} Dan: [{#1}]}}}

\newcommand{\nindep}{\mbox{$\not\!\perp\!\!\!\perp$}}
\newcommand{\indep}{\mbox{$\perp\!\!\!\perp$}}

\newcommand{\satt}{\mathbf{a}}
\newcommand{\att}{\mathbf{A}}
\newcommand{\sx}{\mathbf{x}}
\newcommand{\bx}{\mathbf{X}}

\newcommand{\rwq}{{ {\mc{Q}_{rw}}}}

\newcommand*{\rom}[1]{\expandafter\@slowromancap\romannumeral #1@}
\newcommand{\RNum}[1]{\uppercase\expandafter{\romannumeral #1\relax}}

\newcommand{\algorithmicbreak}{\textbf{Break}}

\newcommand{\mmb}{{\bf MB}}

\newcommand{\mb}[1]{{\mathbf{#1}}}
\newtheorem{defn}{Definition}[section]

\newtheorem{example}[defn]{Example}
\newtheorem{exa}[defn]{Example}

\newtheorem{theo}[defn]{Theorem}
\newtheorem{prop}[defn]{Proposition}

\newtheorem{assumption}[defn]{Assumption}
\newcommand{\proj}[1]{{\Pi}}
\newcommand{\sel}[1]{{\sigma}}

\newcommand{\cut}[1]{}
\newcommand{\eat}[1]{}

\newcommand{\defeq}{\stackrel{\text{def}}{=}}

\def\@settitle{\begin{center}%
		\baselineskip14\p@\relax
		\bfseries
		\uppercasenonmath\@title
		\@title
		\ifx\@subtitle\@empty\else
		\\[1ex]\uppercasenonmath\@subtitle
		\footnotesize\mdseries\@subtitle
		\fi
	\end{center}%
}
\def\subtitle#1{\gdef\@subtitle{#1}}
\def\@subtitle{}
\usepackage{booktabs} 
\usepackage[T1]{fontenc}
\usepackage[utf8]{inputenc}
\usepackage{authblk}

\begin{document}


\title{%
Bias in OLAP Queries: Detection, Explanation, and Removal \footnote{This paper is an extended version of a paper presented at SIGMOD 2018 \cite{salimi2018bias}.} \\[1ex]
	\footnotesize\mdseries
	(think twice about your group-by query) 
}
\author[1]{Babak Salimi \thanks{bsalimi@cs.washington.edu}}
\author[2]{Johannes Gehrke \thanks{johannes@microsoft.com }}
\author[1]{ Dan Suciu \thanks{suciu@cs.washington.edu}}
\vspace{-1cm}
\affil[1]{University of Washington}
\affil[2]{Microsoft}




	%


\date{}


\maketitle
\vspace{-1cm}
\begin{abstract}
On line analytical processing (OLAP) is an essential element of decision-support systems. OLAP tools provide insights and understanding needed for improved decision making. 
However, the answers to OLAP queries can be biased and lead to perplexing and incorrect insights.  In this paper, we propose HypDB, a system to detect, explain, and to resolve bias in decision-support queries. We give a simple definition of a \emph{biased query}, which performs a set of independence tests on the data to detect bias. We propose a novel technique that gives explanations for bias, thus assisting an analyst in understanding what goes on. Additionally,  we develop an automated method for rewriting a biased query into an unbiased query, which shows what the analyst intended to examine. In a thorough evaluation on several real datasets we show both the quality and the performance of our techniques, including the \emph{completely automatic discovery} of the revolutionary insights from a famous 1973 discrimination case. 
\end{abstract}

%
%




\section{Introduction}
\label{sec:intro}

On line analytical processing (OLAP) is an essential element of
decision-support systems.  OLAP tools enable the capability for
complex calculations, analyses, and sophisticated data modeling; this
aims to provide the insights and understanding needed for improved
decision making. Despite the huge progress OLAP research has made in
recent years, the question of whether these tools are truly suitable for
decision making remains unanswered \cite{freitas2006we,
  Binnig2017TowardSI}.  The following example shows how insights
obtained from OLAP queries can be perplexing and lead to poor business
decisions.

\begin{figure*}[t!]

	\hspace*{-.9cm}	\includegraphics[scale=0.51]{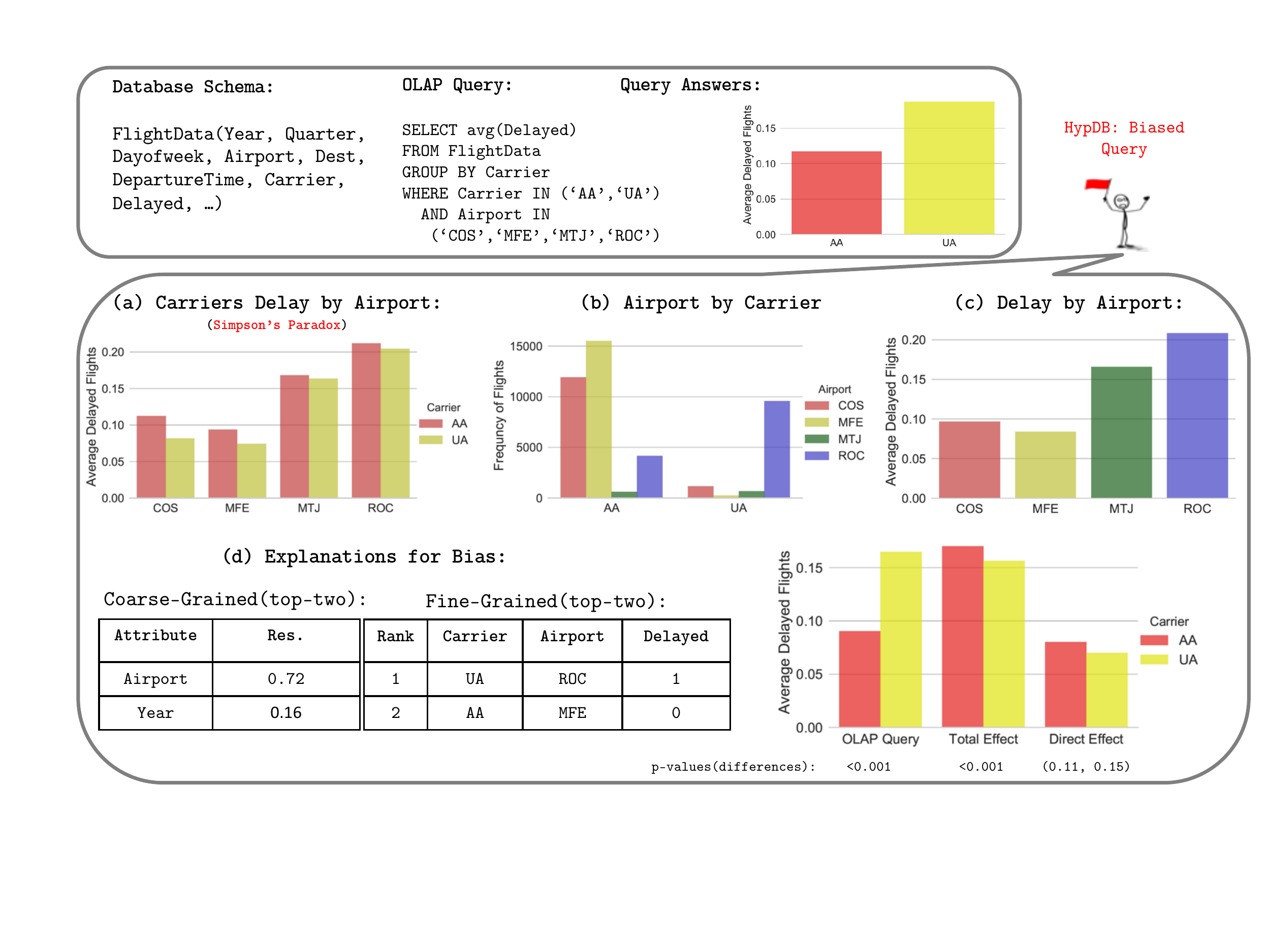}
	\vspace*{-1.5cm}
	\caption{\bf An OLAP query that computes the average of delayed flights for two
		carriers at four airports in Ex.~\ref{ex:simp}. While AA has a lower average delay in the fours airport, UA has a lower average delay
		in each individual airport.	\sys\ explains away
		the anomaly, known as Simpson's paradox,  by inferring confounding attributes and query rewiring.}
	\label{fig:simpex}
\end{figure*}

\begin{exa} \label{ex:simp} Suppose a company wants to choose between
  the business travel programs offered by two carriers, American
  Airlines (AA) and United Airlines (UA).  The company operates at
  four airports: Rochester (ROC), Montrose (MTJ), McAllen Miller (MFE)
  and Colorado Springs (COS). It wants to choose the carrier with the
  lowest rate of delay at these airports.  To make this decision, the
  company's data analyst uses FlightData, the historical flight data
  from the U.S. Department of Transportation (Sec. \ref{sec:setup});
  the analyst runs the group-by query shown in Fig.~\ref{fig:simpex} to
  compare the performance of the carriers.  Based on the analysis
  at the top of Fig.~\ref{fig:simpex}, the analyst recommends choosing AA
  because it has a lower average flight delay.

  Surprisingly, this is a wrong decision.  AA has, in fact, a higher average
  delay than UA at each of the four airports
  Fig. \ref{fig:simpex}(a). This trend reversal, known as Simpson's
  paradox, occurs as a result of  {\em confounding influences}.  The Airport has a confounding
  influence on the distribution of the carriers and departure delays,
  because its distribution differs for AA and for UA
  (Fig.~\ref{fig:simpex} (b) and (c)): AA has many more flights from
  airports that have relatively few delays, like COS and MFE, while UA
  has more flights from ROC, which has relatively many delays.  Thus,
  AA seems to have overall lower delay
  only because it has many flights from airports that in general
  have few delays.  At the heart of the issue is an incorrect interpretation of the query;
  while the analyst's goal is to compare the {\em
    causal effect} of the carriers on delay, the OLAP query 
  measures only their {\em association}.
\end{exa}

A principled business decision should rely on performing a hypothesis
test on the causal effect of choosing between two (or more)
alternatives, $T=t_0$ or $T=t_1$, on some outcome of interest, $Y$.
Data analysts often reach for a simple OLAP query that computes the
average of $Y$ on the two subgroups $T=t_0$ and $T=t_1$, called {\em
  control} and {\em treatment} subpopulations, but, as exemplified in
Fig.~\ref{fig:simpex}, this leads to incorrect decisions.
%
%
\trv{The gold standard for such causal hypothesis testing is a {\em
    randomized experiment} or an {\em A/B test}, called as such
  because the treatments are assigned to subjects randomly.  In contrast,
  business data is {\em observational}, defined as data recorded
  passively and subject to selection bias. }
Although causal inference in observational data has been studied in
statistics for decades, no causal analysis tools exist for OLAP
systems.  Today, most data analysts still reach for the simplest
group-by queries, potentially leading to biased business decisions.

In this paper, we propose \sys, a system to detect, explain, and
resolve bias in decision-support queries.
\srv{
  Our first contribution is a new formal definition of a {\em biased
    query} that enables the system to detect bias in OLAP queries by
  performing a set of independence tests on the data. Next, we
  proposed a novel technique to find \emph{explanations} for the bias and to
  rank these explanations, thus assisting the analyst in understanding
  what goes on.  Third, we describe a query rewriting technique to
  eliminate the bias from queries.  To enable \sys\ to perform these
  types of causal analysis on the data, we develop a novel algorithm
  to compute covariates
  in an efficient way.
  Finally, we perform extensive experiments on several real
  datasets and a set of synthetic datasets, demonstrating that the
  proposed method outperforms the state of the art causal
  discovery methods and that \sys\ can be used interactively to
  detect, explain, and resolve bias at query time.  }

At the core of any causal analysis is the notion of a \emph{covariate}, an attribute that is correlated (``covaries") with the outcome and unaffected by the treatment.
  For example, a
$\text{Group By } T$ query is biased if there exists a set of
covariates $\mb{Z}$ whose distribution differs for the different
groups of $T$; Fig.\ref{fig:simpex}(b) shows that Airport is a
covariate, as its distribution differs for AA and for UA. Thus, the
OLAP query in Fig.\ref{fig:simpex} is biased {w.r.t. Airport}. In order to draw causal conclusions from a biased query, one needs to control
for covariates and thereby to eliminate other possible explanation for a causal connection indicated by the query.
%
%
\srv{The statistics literature typically assumes that the covariates
  are given. An algorithmic approach for finding the covariates was
  developed by Judea Pearl~\cite{de2011covariate,pearl2009causal}, who
  assumes that we are given a {\em causal DAG} over the set of
  attributes, where an edge represents a potential cause-effect
  between attributes, then gives a formal criterion, called the {\em
    back-door criterion}, to identify the covariates.  However, in our
  setting the causal DAG is not given, one has to first compute the
  entire causal DAG from the data in order to apply Pearl's back-door
  criterion.  Notice that a causal DAG does not rank covariates: a
  variable either is, or is not a covariate.
}
Computing the entire causal DAG is inapplicable to interactive OLAP
queries for several reasons.
\srv{First, computing the DAG is expensive, since causal DAG learning
  algorithms perform an exponential (in the number of attributes)
  number of iterations over the data.  It is not possible to
  precompute the causal DAG either, because each OLAP query selects a
  different sub-population through the WHERE condition: this
  phenomenon is called {\em population heterogeneity} in statistics.
  Thus, the causal DAG must be computed at query time.  In addition,
  state of the art causal DAG discovery (CDD) algorithms are not
  robust to sparse subpopulations, which makes them inapplicable for
  OLAP queries with selective WHERE conditions.  We note that one
  should not confuse a causal DAG with an OLAP data cube: they are
  different things.  If a precomputed OLAP cube is available, then the
  computation of the causal DAG can be sped up.  But notice that
  database systems usually limit the data cube to 12 attributes,
  because its size is exponential in the number of attributes; in contrast,
  causal analysis often involves many more attributes. The  FlightData
  dataset in our example has 101 attributes.
}
Second, integrity constraints in databases lead to {\em logical
  dependencies} that totally confuse CDD algorithms: for example,
FlightData satisfies the FDs ``AirportWAC" $\Rightarrow$ Airport'' and therefore conditioning on
AirportWAC (aiport world area code), Airport becomes independent from the rest of the
attributes, which breaks any causal interaction between Airport and
other attributes.  Typically, attributes with high entropy such as ID,
FlightNum, TailNum, etc., participate in some functional dependencies.
Any CDD algorithm must be adapted to handle logical dependencies
before applying it to OLAP data.


In order to incorporate causal inference in OLAP, \sys\ makes two key
technical contributions.  First, we propose a novel method for
covariate discovery that does not compute the entire causal DAG, but explores
only the subset relevant to the current query
 We empirically show that our
method is competitive with state of the art CDD algorithms, yet it
scales well with large and high-dimensional data, is robust to sparse
subpopulations, and can handle functional dependencies on the fly.
Second, we propose a powerful optimization to significantly speed up
the {\em Monte Carlo permutation-test}, which is an accurate, but
computationally expensive independence test needed throughout \sys\
(detecting biased queries, explaining the bias, and resolving it by
query rewriting).
Our optimization consists of generating permutation
samples without random shuffling of data, by sampling from contingency
tables and conditioning groups.

\srv{Finally, a key novelty of \sys\ is the ability to explain its
  findings, and rank the explanations. We introduce novel definitions
  for fine-grained and coarse-grained explanations of a query's bias
  (Example~\ref{ex:sx}). We empirically show that these
  explanations are crucial for decision making and reveal illuminating
  insights about the domain and the data collection process.  For
  instance, \sys\ reveals an inconsistency in the adult dataset
  \cite{adult} (Section~\ref{sec:experi}).}

\vspace{-0.1cm}
\begin{exa}  \label{ex:sx}
  \sys\ will detect that the query in Fig.~\ref{fig:simpex} is biased
  and will explain the bias by computing a list of covariates ranked by
  their {\em responsibility}.  Fig.~\ref{fig:simpex} (d) shows that
  Airport has the highest responsibility, followed by Year; this
  provides valuable information to the data analyst for understanding
  the trend reversal.  Finally, \sys\ rewrites the original query into  a query of the form Listing~\ref{rfq} in
  order to compute both the {\em total effect} and the {\em direct
    effect}. The total effect measures the expected changes in the delay when the carrier is set to AA and UA by a
  hypothetical external intervention. The direct effect measures the effect that is not
  mediated by other variables, such as destination and late arrival
  delay.  Fig.~\ref{fig:simpex} (d) shows that UA has a slightly
  better performance than AA in terms of total effect, but there is no
  significant difference for the direct effect.
\end{exa}

An important application of \sys, which we demonstrate in the
empirical evaluation, is to detect {\em algorithmic
  unfairness}~\cite{pedreshi2008discrimination,zemel2013learning,tramer2017fairtest}.
Here, the desire is to ensure that a machine learning algorithm does
not make decisions based on protected attributes such as gender or
race, for example in hiring decisions.  While no generally accepted
definition exists for algorithmic unfairness, it is known that any
valid proof of unfairness requires evidence of causality
\cite{foster2004causation}. For example, in gender discrimination, the
question is whether gender has any direct effect on income or hiring
\cite{pearl2001direct}.  We show how to use \sys\ post factum to
detect unfairness, by running a group-by query on the protected
attribute and checking for biasness.  We detect algorithmic unfairness
and obtain insights that go beyond state of the art tools such as
FairTest \cite{tramer2017fairtest}.

The main contributions of this paper are as follows: We provide a formal
definition of a biased query based on independence tests in the data
(Sec.~\ref{sec:dbias}); give a definition of responsibility and
contribution, allowing us to rank attributes and their ground levels
by how well they explain the bias of a query (Sec.~\ref{sec:ebias});
and describe a query rewriting procedure that eliminates bias from
queries (Sec.~\ref{sec:rbias}).  Then, we propose a novel algorithm
for computing covariates without having to compute the complete causal
DAG (Sec.~\ref{sec:acd}).  Next, we describe optimization techniques
that speed up the independence test based on the Monte Carlo
permutation-test (Sec.~\ref{sec:cit}). We propose some optimizations
to speed up all components of our system
(Sec.~\ref{sec:mbd}). Finally, perform an extensive evaluation using
four real datasets and a set of synthetic datasets
(Sec.~\ref{sec:experi}).

\vspace{-.1cm}


\section{Background and Assumptions}
\label{sec:peri}

We fix a relational schema with attributes $\att=\{X_1,\ldots,X_k\}$
and discrete domains $Dom(X_i), i=1,k$.  We use lower case letters to
denote values in the domains, $x \in Dom(X)$, and use bolded letters
for sets of attributes $\mb X$, or tuples $\sx \in Dom(\bx)$
($\defeq \prod_{X \in \bx} Dom(X)$) respectively.  A {\em database
  instance} is a set of tuples with attributes $\att$.  We denote its
cardinality by $n$.

 \vspace*{-0.2cm}
{\small
\begin{lstlisting}[language=SQL,escapechar=@,language=SQL, escapeinside={*}{*},
basicstyle=\ttfamily,frame=tlrb,caption=An OLAP query $\mc{Q}$.,label=olap]{}
SELECT *$T$*,*$\mb{X}$*,avg(*$Y_1$*), ... ,avg(*$Y_e$*)
FROM *$D$*
WHERE *$\mb C$*
GROUP BY *$T$*,*$\mb{X}$*
\end{lstlisting}
}

%
\frv{We restrict the OLAP queries to group-by-average queries, as
  shown in Listing~\ref{olap}.  We do not consider more complex OLAP
  queries (drill-down, roll-up or cube queries); we also do not
  consider aggregate operators other than average because they are not
  needed in causal analysis.}
To simplify the exposition we assume $T$ and $Y_i$ take only two
values, $Dom(T)=\{t_0,t_1\}$, $Dom(Y_i)=\{0,1\}$ and denote
$Dom(\mb{X})=\{\mb{x}_1 \ldots \mb{x}_m\}$.  For each $i=1,m$ we call
the condition $\Gamma_i \defeq \mb{C} \land (\mb{X}=\mb{x}_i)$ a {\em
  context} for the query $\mc{Q}$.  We interpret the query as follows:
for each context, we want to compute the difference between
$\text{avg}(Y_i)$ for $T=t_1$, and for $T=t_0$.

\trv{We make the following key assumption, standard in statistics.}  The
database $D$ is a uniform sample from a large population (e.g., all
flights in the United States, all customers, etc.), obtained according
to some unknown distribution $\pr(\att)$.  Then, the query $\mc{Q}$
represents a set of estimates $\E[(Y_1,\dots,Y_e)|T=t_j,\Gamma_i]$.

We assume familiarity with the notions of entropy $H(\bx)$,
conditional entropy $H(\mb{Y}|\bx)$, and mutual information
$I(\bx;\mb{Y}|\mb{Z})$ associated to the probability distribution
$\pr$; see the Appendix for a brief review.  Since $\pr$ is not known,
we instead estimate the entropy from the sample $D$ using the
Miller-Madow estimator \cite{miller1955note}.

\paragraph*{\bf The Neyman-Rubin Causal Model (NRCM)} \sys\ is based on
the Neyman-Rubin Causal Model
(NRCM)~\cite{holland1986statistics,rubin1970thesis}, whose goal is to
study the causal effect of a {\em treatment} variable
$T \in \set{t_0, t_1}$ on an {\em outcome} variable $Y$.  The model
assumes that every object (called {\em unit}) in the population has
{\em two} attributes, $Y(t_0), Y(t_1)$, representing the outcome both
when we don't apply, and when we do apply the treatment to the unit.
The {\em average treatment effect (ATE)} of $T$ on $Y$ is defined as:
\begin{align}
  \ate(T,Y) \defeq \E[Y(t_1)-Y(t_0)] = \E[Y(t_1)] - \E[Y(t_0)]  \label{eq:ate}
\end{align}
In general, ATE cannot be estimated from the data, because for each
unit one of $Y(t_0)$ or $Y(t_1)$ is missing; this is called the {\em
  fundamental problem of causal inference}~\cite{Holland1986}.  To
compute the ATE, the statistics literature considers one of two
assumptions.  The first is the {\em independence assumption}, stating
that $(Y(t_0),Y(t_1))$ is independent of $T$.  Independence holds only
in randomized data, where the treatment $T$ is chosen randomly, but
fails in observational data, which is the focus of our work; we do not
assume independence in this paper, but, for completeness, include its
definition in the appendix.  For observational data, we consider a
second, weaker assumption~\cite{rubin1986statistics}:

\begin{assumption} \label{assumption:1} The data contains a set of
  attributes $\mb Z \subseteq \att$, called {\em covariates},
  satisfying the following property: forall $\mb{z} \in Dom(\mb Z)$,
  (1) $({Y(t_0), Y(t_1) \bigCI T} | \mb{Z}=\mb{z})$ (this is called
  {\em Unconfoundedness}), and (2)
  $0 < \Pr(T = t_1 | \mb{Z}=\mb{z}) < 1$ (this is called {\em
    Overlap}).
\end{assumption}

Under this assumption, $\ate$ can be computed using the following
adjustment formula:
%
\begin{eqnarray} \small
  \ate(T,Y) =  \sum_{\mb{z} \in Dom(\mb{Z})} (\E[Y|T=t_1,\mb{z}]-\E[Y|T=t_0,\mb{z}]) \ \Pr(\mb{z}) \label{eq:adj}
\end{eqnarray}

\begin{example}
  Referring to our example~\ref{ex:simp}, we assume that each 
  flight has {\em two} delay attributes, $Y(\text{AA})$ and
  $Y(\text{UA})$, representing the delay if the flight were serviced
  by AA, or by UA respectively.  Of course, each flight was operated
  by either AA or UA, hence $Y$ is either
  $Y(\text{AA})$ or $Y(\text{UA})$ in the database; the other one is missing, and
  we can only imagine it in an alternative, counterfactual world.  The
  independence assumption would require a controlled experiment, where
  we assign randomly each flight to either AA or UA, presumably right
  before the flight happens, clearly an impossible task.  Instead,
  under the Unconfoundedness assumption we have to find sufficient
  covariates, such that, after conditioning, both delay variables are
  independent of which airline operates the flight.  We note that
  ``independence'' here is quite subtle.  Clearly the {\em delay} $Y$
  is dependent of the {\em airline}, because it depends on whether the
  flight is operated by AA or UA.  What the assumption states is that,
  after conditioning, the {\em delay of the flight if it were operated
    by AA} (i.e. $Y(\text{AA})$) is independent of whether the flight
  is actually operated by AA or UA, and similarly for $Y(\text{UA})$.
\end{example}

To summarize, in order to compute $\ate$, one has to find the
covariates $\mb Z$.  The {\em overlap} condition is required to ensure
that \eqref{eq:adj} is well defined, but in practice overlap often
fails on the data $D$.  The common approach is to select covariates
$\mb Z$ that satisfy only Unconfoundedness, then estimate
\eqref{eq:adj} on the subset of the data where overlap holds, see
Sec.~\ref{sec:rbias}.  Thus, our goal is to find covariates satisfying
Unconfoundedness.

{\bf Causal DAGs.} A {\em causal DAG} $G$ is graph whose
nodes are the set of attributes $V(G)=\att$ and whose edges $E(G)$
capture all potential causes between the
variables~\cite{pearl2001direct,pearl2009causality,de2011covariate}.
We denote $\mb{PA}_X$ the set of parents of $X$.  If there exists a
directed path from $X$ to $Y$ then we say that $X$ is an {\em
  ancestor} or a {\em cause} of $Y$, and say that $Y$ is a {\em
  descendant} or an {\em effect} of $X$.  A probability distribution
$\Pr(\mb{A})$ is called {\em causal}, or {\em DAG-isomorphic}, if
there exists a DAG $G$ with nodes $\mb{A}$ that captures precisely its
independence relations\cite{pearl2014probabilistic,
  pearl2009causality,spirtes2000causation}: the formal definition is
in the appendix, and is not critical for the rest of the paper.
Throughout this paper we assume $\Pr$ is DAG-isomorphic.

{\bf Covariate Selection.}
Fix some database $D$, representing a sample of some unknown
distribution $\Pr$, and suppose we want to compute the causal effect
of an attribute $T$ on some other attribute $Y$.  Suppose that we
computed somehow a causal DAG $G$ isomorphic to $\Pr(\mb{A})$.
Pearl~\cite{pearl1993bayesian} showed that the parents of $T$ are
always a sufficient set of covariates, more precisely:
%
%
%
\begin{prop}~\cite[Th. 3.2.5]{pearl2009causality} \label{prop:par_cov}
  Fix two attributes $T$ and $Y$.  Then the set of parents,
  $\mb Z \defeq \mb{PA}_T$ satisfies the Unconfoundedness property.
\end{prop}
In \sys\ we always choose $\mb{PA}_T$ as covariates, and estimate
$\ate$ using Eq.~\eqref{eq:adj} on the subset of the data where
overlap holds; we give the details in Sec.~\ref{sec:rbias}.

{\bf Learning the Parents from the Data.}
The problem is that we do not have the causal dag $G$, we only have
the data $D$.  Learning the causal DAG from the data is considered to
be a challenging task, and there are two general approaches.  The {\em
  score-based} approach \cite{heckerman1998tutorial} uses a heuristic
score function on DAGs and a greedy search. The {\em constraint-based}
approach~\cite{Spirtes:book01,pearl2010introduction} builds the graph
by repeatedly checking for independence relations in the data.  Both
approaches are expensive, as we explain in Sec.~\ref{sec:acd}, and
unsuitable for interactive settings.  Furthermore, in our application
the causal DAG must be computed at query time, because it depends on
the WHERE condition of the query.  Instead, to improve the efficiency
of \sys, we compute only $\mb{PA}_T$, by using the Markov Boundary.


\vspace{-.1cm}
\begin{defn} \cite{pearl2014probabilistic} Fix a probability
  distribution $\Pr(\mb A)$ and a variable $X \in \mb A$.  A set of
  variables $\mb B \subseteq \mb A - \set{X}$ is called a {\em Markov
    Blanket} of $X$ if $(X \indep \mb A - \mb B - \set{X} | \mb B)$;
  it is called a {\em Markov Boundary} if it is minimal w.r.t. set
  inclusion.
\end{defn}
Next, we relate the Markov boundary of $T$ with $\mb{PA}_T$:
\begin{prop}
  \cite[The. 2.14]{neapolitan2004learning} \label{prop:mbdag} Suppose
  $P(\att)$ is DAG-isomorphic with $G$. Then for each variable $X$,
  the set of all parents of $X$, children of $X$, and parents of
  children of $X$ is the unique Markov boundary of $X$, denoted
  $\mb B(X)$.
\end{prop}

Thus, $\mb{PA}_T \subseteq \mb B(T)$.  Several algorithms exists in
the literature for computing the blanket, $\mb B(T)$, for example the
Grow-Shrink~\cite{margaritis2000bayesian}.  In Sec.~\ref{sec:acd} we
describe a novel technique that, once $\mb B(T)$ is computed, extracts
the parents $\mb{PA}_T$.

{\bf Total and Direct Effects.} $\ate$ measures the total
effect of $T$ on $Y$, aggregating over all directed paths from $T$ to
$Y$.  In some cases we want to investigate the {\em natural direct
  effect}, $\nde$~\cite{pearl2001direct}, which measures the effect
only through a single edge from $T$ to $Y$.  Its definition is rather
technical and deferred to the appendix.  For the rest of the paper it
suffices to note that it can be computed using the following
formula~\cite{pearl2001direct}, $\nde(T,Y)=$
\begin{align}
 \sum_{\parbox{5mm}{\mbox{\scriptsize $(\mb{z},\mb{m}) \in Dom(\mb{Z},\mb{M})$}}}(\E[Y|T=t_0, \mb{m} ]-
  \E[Y|T=t_1,\mb{m}])\Pr(\mb{m}| T=t_1,\mb{z} )\Pr(\mb{z}) \  \label{eq:medi}
\end{align}
where the covariates are $\mb{Z} \defeq \mb{PA}_T$, and
$\mb{M} \defeq \mb{PA}_Y -\{T\}$ is called the set of {\em mediators}.
\sys\ computes both total and direct effect.

\vspace{-.1cm}
\section{Biased OLAP Queries}
\label{sec:unfq}

We introduce now the three main components of \sys: the definition of
bias in queries, a novel approach to explain bias, and a technique to
eliminate bias.  Throughout this section, we will assume that the user
issues a query $\mc{Q}$ (Listing~\ref{olap}), with intent to study the
causal effect (total or direct) from $T$ to $Y_j$, for each outcome
variable $Y_j$, and for each context $\Gamma_i$.  We assume that we
are given the set of covariates $\mb Z$ (for the total effect) or
covariates $\mb Z$ and mediators $\mb M$ (for the direct effect); next
section explains how to compute them.  We
assume that all query variables other than treatment and outcome are
included in $\mb Z$, or $\mb Z \cup \mb M$ respectively.

\vspace*{-0.2cm}
\subsection{Detecting Bias} \label{sec:dbias}

Let $\mb{V}$ denote $\mb{Z}$ for total effect, or $\mb{Z} \cup \mb{M}$
for direct effect.

\begin{defn} \label{def:unfair} We say the query $\mc{Q}$ is balanced
  w.r.t. a set of variables $\mb{V}$ in a context $\Gamma_i$ if the
  marginal distributions $\Pr(\mb{V}|T=t_0, \Gamma_i)$ and
  $\Pr(\mb{V}|T=t_1, \Gamma_i)$ are the same.
\end{defn}

Equivalently, $\mc{Q}$ is balanced w.r.t. $\mb{V}$ in the context
$\Gamma_i$ iff $(T \indep \mb{V} | \Gamma_i)$.  We prove (in the appendix):




\begin{prop} \label{def:biasedq} Fix a context $\Gamma_i$ of the query
  $\mc{Q}$, and let $\Delta_i$ denote the difference between
  $\text{avg}(\mb Y)$ for $T=t_1$ and for $T=t_0$ ($\Delta_i$
  estimates
  $\E[\mb{Y} |T=t_1,\Gamma_i]-\E[\mb{Y} |T=t_0|\Gamma_i]$). Then:
	\begin{itemize}
        \item[(a)] if $\mc{Q}$ is balanced w.r.t. the covariates
          $\mb Z$ in the context $\Gamma_i$, then $\Delta_i$ is an
          unbiased estimate of the $\ate$ (Eq.~\eqref{eq:ate}).  In
          this case, with some abuse, we say that $\mc{Q}$ is {\em
            unbiased for estimating total effect}.
        \item[(b)] if $\mc{Q}$ is balanced w.r.t. the covariates and
          mediators $\mb Z \cup \mb M$ in the context $\Gamma_i$, then
          $\Delta_i$ is an unbiased estimate of $\nde$
          (Eq.~\eqref{eq:nde}); we say that $\mc{Q}$ is {\em unbiased
            for estimating direct effect}.
	\end{itemize}
\end{prop}


In other words, if the query is unbiased w.r.t. the covariates $\mb Z$
(and mediators $\mb M$) then the user's query is an unbiased
estimator, as she expected.  In that case the groups are comparable in
every relevant respect, i.e., the distribution of potential covariates
such as age, proportion of male/female, qualifications, motivation,
experience, abilities, etc., are similar in all groups.  The population
defined by the query's context behaves like a randomization
experiment.

During typical data exploration, however, queries are biased.  In
Ex.~\ref{ex:simp}, the distribution of the covariate Airport differs
across the two groups formed by Carrier
(Fig.~\ref{fig:simpex}(b)). This makes the groups incomparable and the
query biased in favor of AA, because AA has many more flights from
airports that have few delays, like COS and MFE, while UA has more
flights from ROC, which has many delays
(Fig.~\ref{fig:simpex} (b) and (c)).


To detect whether a query is unbiased w.r.t. $\mb{V}$, we need to
check if $(T \indep \mb{V} | \Gamma_i)$, or equivalently if
$I(T; \mb{V} | \Gamma_i) = 0$, where $I$ is the conditional mutual
information.
\trv{Recall that $I$ is defined by the unknown probability
  distribution $\Pr$ and cannot be computed exactly.  Instead, we have
  the database $D$, which is a sample of the entire population, we
  estimate $\hat I$, then check dependence by rejecting the null
  hypothesis $ I(T; \mb{V} | \Gamma_i) = 0$ if the $p$-value is
  small enough. (Even if $D$ consists of the entire population, the
  exact equality $I(T; \mb{V} | \Gamma_i) = 0$ is unlikely to hold for
  the uniform distribution $\Pr$ over $D$.) We discuss this in detail
  in Sec.~\ref{sec:cit}.
}
%
%
%
%
For an illustration, in Ex.~\ref{ex:simp},
$\hat{I}(\text{Carrier}; \text{Aiport} | \Gamma)=0.25\neq 0$ with
p-value$<0.001$, where $\Gamma$ is the context of the four
airports. Thus, the query is biased.


\ignore{
Note that Def. \ref{def:unfair} concerns only the total effects $T$ on
$Y$. However, it can be simply extended to define biased queries
w.r.t. direct effects.  In this case we assume a set of potential mediating variables $\mb{M} \subseteq \att- (\mb{Z} \cup  \{T\})$ are also given. \dan{please give the definition, since it is one
  of the main contributions of the paper}

\begin{defn} \label{def:unfair} We say the query $\mc{Q}$ is unbiased
  for computing direct effects w.r.t. $\mb{Z} \cup \mb{M}$ in the
  context $\Gamma_i$ if the distribution of $\mb{Z} \cup \mb{M}$ is
  similar in the two groups, i.e.,
  $P(\mb{Z} \cup \mb{M}|T=t_0, \Gamma_i)$ is equal to
  $P(\mb{Z} \cup \mb{M}|T=t_1,\Gamma_i)$; otherwise, we say $\mc{Q}$
  is biased for computing direct effects w.r.t. $\mb{Z} \cup
  \mb{M}$.  \end{defn}

In a similar manner, to test whether a query is biased, one can check if
$I(T; \mb{Z} \cup \mb{M} | \Gamma_i) = 0$.
}
\vspace*{-0.2cm}
\subsection{Explaining Bias} \label{sec:ebias}

In this section, we propose novel techniques to explain the bias in
the query $\mc{Q}$.  We provide two kinds of explanations: {\em coarse
  grained} explanations consist of a list of attributes $Z \in \mb{Z}$
(or $\mb Z \cup \mb{M}$), ranked by their responsibility to the bias,
and {\em fine grained} explanations, consisting of categories (data values) of each  attribute $Z$, ranked by their contribution to bias.


{\bf Coarse-grained.} Our coarse-grained
explanation consists of ranking the variables $\mb{V}$ (which is
either $\mb{Z}$ or $\mb{Z} \cup \mb{M}$), in terms of their
responsibilities for the bias, which we measure as follows:

\begin{defn} (Degree of Responsibility): \label{def:res} We define the
  {\em degree of responsibility} of a variable $Z \in \mb{V}$ in the
  context $\Gamma_i$ as
  \begin{eqnarray}
    \rho_Z= \frac{I(T;\mb{V}|\Gamma_i)-I(T;\mb{V}|Z,\Gamma_i)}{ \sum_{V \in \mb V}  I(T;\mb{V}|\Gamma_i)-I(T;\mb{V}|V,\Gamma_i) } \label{eq:res}
  \end{eqnarray}
\end{defn}

When $Z \in \mb V$, the quantity
$I(T;\mb{V}|\Gamma_i)-I(T;\mb{V}|Z,\Gamma_i)$ is
always\footnote{Dropping the context, we have
  $I(T;\mb{V}) - I(T;\mb{V}|Z) = (H(T)+H(\mb V)-H(T\mb V)) -
  (H(TZ)+H(\mb{V})-H(T\mb V)-H(Z)) = H(T)+H(Z)-H(TZ) \geq 0$ by
  submodularity. Notice that, if $Z \not\in \mb{V}$, then it is known
  that this difference may be $< 0$.}  $\geq 0$.  Therefore $\rho_Z$
is simply the normalized value of some positive quantities, and thus
$0 \leq \rho_{Z} \leq 1$.  The larger $\rho_{Z}$, the more responsible
$V$ is for bias. The intuition is that there is no bias iff
$I(T;\mb{V}|\Gamma_i)=0$, thus the degree of responsibility measures
the contribution of a single variable to the inequality
$I(T;\mb{V}|\Gamma_i)> 0$.
%
%




\sys\ generates coarse-grained explanations for a biased query, by
ranking covariates and mediators in terms of their responsibilities.
In Ex.~\ref{ex:simp}, the covariates consists of attributes such as
Airport, Day, Month, Quarter, Year.  Among them Airport has the
highest responsibility, followed by Year (Fig.~\ref{fig:simpex}
(d)). See Sec.~\ref{sec:experi} for more examples.


{\bf Fine-Grained.} A fine-grained explanation for
a variable $Z \in \mb{V}$ is a triple $(t,y,z)$, where $t \in Dom(T)$,
$y \in Dom(Y)$, $z \in Dom(Z)$, that highly contributes to both
$I(T;Z)$ and $I(Y;Z)$. These triples explain the confounding (or
mediating) relationships between the ground levels.  We measure these
contributions as follows:

\begin{defn} (Degree of contribution): Given two variables
  $X,Y \in \att$ with $ I(X;Y)>0$ and a pair $(x, y) \in Dom(XY)$, we
  define the degree of contribution of $(x, y)$ to $I(X;Y)$ as:
	\begin{eqnarray}
	\kappa_{(x,y)}=  {\pr(x,y) \log(\frac{\pr(x,y)}{\pr(x)\pr(y)})} \label{eq:cntri}
	\end{eqnarray}
\end{defn}

Mutual information satisfies
$I(X;Y) = \sum_{(x,y)\in Dom(\mb X, \mb Y)} \kappa_{(x,y)}$.  Thus, a
pair $(x, y)$ can either make a {\em positive} ($\kappa_{(x, y)}>0$),
{\em negative} ($\kappa_{(x, y)}<0$), or {\em no contribution}
($\kappa_{(x, y)}=0$) to $I(X;Y)$.

\ignore{
we rank triples $(z, t, y)$ for $z \in Dom(Z)$, $t \in Dom(T)$ and $y \in Dom(Y)$,  in terms of their contribution
to the association between $Z$ and $T$; and $Z$ and $Y$. Intuitively, we are interested in triples that highly contribute
to {\em both} associations\ignore{ In this direction, based on the fact that  $I(T;Z)=\sum_{t \in T} \sum_{z \in Z} p(t,z) \log(\frac{p(t,z)}{p(t)p(z)} )$}.
}

To compute the contribution of the triples $(t,y,z)$ to both
$I(T;Z|\Gamma_i)$ and $I(Y;Z|\Gamma_i)$ and generate explanations,
\sys\ proceeds as follows.  It first ranks all triples
$(t, y, z) \in \Pi_{TYZ}(\sigma_{\Gamma_i}(D))$, based on their
contributions to $\hat{I}(T;Z)$, then it ranks them again based on
their contribution to $\hat{I}(Y;Z)$, then aggregates the two rankings
using Borda's methods~\cite{lin2010rank}; we give the details in the
algorithm \fge\  in Alg.~\ref{alg:fgealg} in the appendix.  \sys\ returns
the top $k$ highest ranked triples to the user.
%
%
%
For example, Fig.~\ref{fig:simpex}(d) shows the triple (Airport=ROC,
Carrier=UA, Delayed=1) as the top explanation for the
$Z = \text{Airport}$ covariate; notice that this captures exactly the
intuition described in Ex.~\ref{ex:simp} for the trend reversal.



\begin{figure}
	{		\small
		\begin{lstlisting}[language=SQL,escapechar=@,language=SQL, escapeinside={|}{|},
		basicstyle=\ttfamily,caption=Refined OLAP query $\rwq$.,label=rfq]
		WITH Blocks
		AS(
		SELECT |$T$|,|$\mb{X}$|,|$\mb{Z}$|,avg(|$Y_1$|) AS Avg1,...,avg(|$Y_e$|) AS Avge
		FROM |$D$|
		WHERE |$\mb C$|
		GROUP BY  |$T$,$\mb{Z}$,$\mb{X}$|),
		Weights
		AS(
		SELECT|$\mb{X}$|,|$\mb{Z},$|count(*)/|$n$| AS W
		FROM |$D$|
		WHERE |$\mb C$|
		GROUP BY  |$\mb{Z}$,$\mb{X}$|
		HAVING count(DISTINCT |$T$|)=2)
		SELECT |$T$|,|$\mb{X}$|,sum(Avg1 * W),...,sum(Avge * W)
		FROM Blocks,Weights
		WHERE Blocks.|$\mb{Z}$| = Weights.|$\mb{Z}$| AND
		Blocks.|$\mb{X}$| = Weights.|$\mb{X}$|
		GROUP BY |$T$|,|$\mb{X}$|
		\end{lstlisting}}
\end{figure}

\vspace*{-0.23cm}
\subsection{Resolving Bias}
\label{sec:rbias}



Finally, \sys\ can automatically rewrite the query to remove the bias,
by conditioning on the covariates $\mb Z$ (recall that we assumed in
this section the covariates to be known).  Listing~\ref{rfq} shows the
general form of the rewritten query $\rwq$ for computing the total
effect of $\mc{Q}$ (Listing~\ref{olap}).
%
%
The query $\rwq$ essentially implements the adjustment formula
Eq.~\eqref{eq:adj}; it partitions the data into blocks that are
homogeneous on $\mb{Z}$. It then computes the average of each
$ Y \in \mb{Y}$ Group by $T,\mb X$, in each block.  Finally, it
aggregates the block's averages by taking their weighted average,
where the weights are probabilities of the blocks.  In order to
enforce the {\em overlap} requirement (Assumption~\ref{assumption:1})
we discard all blocks that do not have at least one tuple with $T=t_1$
and one tuple with $T=t_0$; this pruning technique is used in causal
inference in statistics, and is known as {\em exact
  matching}~\cite{iacus2009cem}.  We express exact matching in SQL
using the condition $\texttt{count(DISTINCT T)}=2$, ensuring that for
every group $t_0, \mb{x}, \texttt{avg}_1, \ldots$ in the query answer,
there exits a matching group $t_1, \mb{x}, \texttt{avg}_1', \ldots$,
and vice versa.  Note that probabilities need to be computed wrt. the size of renaming data after pruning. The API of \sys\ finds these matching groups, and
computes the differences $\texttt{avg}_i'-\texttt{avg}_i$, for
$i=1,e$; this represents precisely the $\ate$ for that context,
Eq.~\eqref{eq:ate}. \frv{The rewritten query associated to Ex.~\ref{ex:simp} is shown in Listing~\ref{rfqex1}, in the appendix.}

\sys\ performs a similar rewriting for computing the direct effects of
$T$ on covariates $\mb{Y}$ and mediators $\mb M$, by implementing the
mediator formula (Eq.~\ref{eq:medi}).

\ignore{
\paragraph*{Significance testing} Let $\hat{a}_{ij}$ and $\hat{a}'_{ij}$ for $j \in \{0,1\}$, respectively be the answers to the queries
$\mc{Q}$ and $\rwq$ (Li. \ref{olap} and \ref{rfq}),	in a context $\Gamma_i$. Recall from Sec.  \ref{sec:peri} that  $\hat{a}_{ij}$ and $\hat{a}'_{ij}$   estimate conditional expectations $a_{ij}$ and ${a}'_{ij}$, respectively.
For instance, $\hat{a}_{ij}$ estimates  $a_{ij}=\E[Y|T=t_j, \Gamma_i]$. Suppose the null-hypothesis $a_{i0}={a}_{i1}$ holds, the problem of testing the significance of an OLAP query result is to measure
how likely it is that to obtain an answer with $\abs{\hat{a}_{i0}-\hat{a}_{i1}}>0$, under the null-hypothesis. Next proposition enables \sys\ to use significant tests for CMI (will be discussed in Sec.  \ref{sec:cit}) to address this problem for $\mc{Q}$ and $\rwq$.

\begin{prop}  \label{obs:gbg_mi}
	Given a context $\Gamma_i$, it holds that: $a_{i0}=a_{i1}$ iff $I(T;Y|\Gamma_i)=0$; and $a'_{i0}=a'_{i1}$ iff $I(T;Y|\mb{Z})$.
\end{prop}
}

\vspace*{-0.2cm}
\section{Automatic Covariates Discovery}
\label{sec:acd}

In this section we present our algorithm for automatic
covariates discovery from the data.  More precisely, given a treatment
variable $T$, our algorithm computes its parents in the causal DAG,
$\mb{PA}_T$, and sets $\mb{Z} = \mb{PA}_T$ (Prop.~\ref{prop:par_cov});
importantly, our algorithm discovers $\mb{PA}_T$ directly from the
data, without computing the entire DAG.


In this section we assume to have an oracle for testing
conditional independence $(U \indep V | W)$ in the data; we describe
this algorithm in the next section.  Using repeated independence
tests, we can compute the Markov Boundary of $T$, $\mmb(T)$,
e.g. using the Grow-Shrink algorithm \cite{margaritis2000bayesian}.
While $\mb{PA}_T \subseteq \mmb(T)$, identifying the parents is
difficult because it is sometimes impossible to determine the
direction of the edges.  For example, consider a dataset with three
attributes $T, W, Z$ and a single independence relation,
$(Z \indep W | T)$.  This is consistent with three causal DAGs:
$Z \rightarrow T \rightarrow W$, or $Z \leftarrow T \leftarrow W$, or
$Z \leftarrow T \rightarrow W$, and $\mb{PA}_T$ is either $Z$, or $W$,
or $\emptyset$.  A {\em Markov equivalence} class
\cite{spirtes2000causation} is a set of causal DAGs that encode the
same independence assumptions, and it is well known that one cannot
distinguish between them using only the data.  For that reason, in
\sys\ we make the following assumption: for every $U \in \mb{PA}_T$
there exists $V \in \mb{PA}_T$ such that $U,V$ are not neighbors.
Intuitively, the assumption says that $\mb{PA}_T$ is big enough: if
$\mb{PA}_T= \emptyset$, then there is no need to choose any covariates
(i.e. $\mb{Z}=\emptyset$), so the only setting where our assumption
fails is when $T$ has a single parent, \frv{or all its parents are
neighbors. In the former case, \sys\ sets $\mb Z=\mmb(T)-\{Y\}$. In the latter case, parents of $T$  can not be learned from data. However, one can compute a set of 
potential parents of $T$ and use them to establish a bound on causal effect. For example, in the causal DAG $G$ consists of edges $(U,V)$, $(V,T)$, $(U,T)$, $(T,Y)$ and $(U,Y)$,
parents of $T$ are neighbors and can not be learned from data. However, to compute the effect of $T$ on $Y$, one can 
learn $\mmb(T)=\{U,V,Y\}$ from data, and then set $\mb Z=\{U,V\}$, $\mb Z=\{U\}$, $\mb Z=\{V\}$ and $\mb Z=\emptyset$, i.e., all subsets of $\mmb(T)-\{Y\}$, to infer a bound on the effect. We leave this extension for future work.} Given our assumption, we prove:




\begin{prop} \label{prop:detec} Let $\Pr$ be DAG-isomorphic with $G$,
  $T\in V(G)$, and $Z \in \mmb(T)$.  Then $Z \in \mb{PA}_T$ iff both
  conditions hold:
  \begin{itemize}
  	\item[(a)] There exists  $W \in \mmb(T)-\set{Z}$ and $\mb{S} \subseteq \mmb(Z)-\set{W,T}$  such that 
       $(Z \indep W | \mb{S}) \land (Z \nindep W | \mb{S}\cup \{T\})$
  	\item[(b)] Forall  $\mb{S'} \subset \mmb(T)-\{Z\}$,  $(Z  \nindep T | \mb{S'})$
  \end{itemize}

\end{prop}

\begin{figure}  \center
	\includegraphics[scale=0.3]{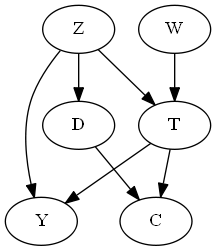}
	\caption{\bf{Example of a causal DAG.}}
		\label{fig:dagex}
	\end{figure}

The intuition behind this proposition is that a necessary condition
for $Z,W \in \mmb(T)$ to be the parents of $T$ is that $T$ be a
{\em collider} in a path between them. \trv{ In causal DAG terms, a common descendant of two nodes is called a collider node, because two arrowheads collide at this node (see Appendix \ref{sec:app_ab} for a formal definition and example).} If $Z$ and $W$ are not neighbors this
can be detected from data by performing a series of independence tests
to check for the condition (a).  For instance, in the causal DAG in
Fig.~\ref{fig:dagex} ($Z \indep W$) but ($Z \not \indep W | T$), thus
(a) holds for $\mb{S}=\emptyset$.  However, (a) is only necessary but
not sufficient. In Fig.~\ref{fig:dagex}, $D \indep W$ and
$D \not \indep W| T$, but $D$ is not a parent of $T$. This would
happen if $T$ were a collider in a path between one of its parents,
e.g., $W$, and a parent of its children, e.g., $D$, that are
(conditionally) independent.\footnote{Note that $T$ can be a collider
  in a path between any pairs of its parents, children, and parents of
  its children, e.g., $(W,C)$, $(W,Y)$ in
  Fig.~\ref{fig:dagex}. However, only two parents or one parent and a
  parent of a child can satisfy (a).} Condition (b) excludes such
cases by removing all those that are not neighbor with $T$.
Furthermore, to check (a) and (b), it is sufficient to restrict the
search to subsets of the relevant Markov boundaries.

 The \pdal\ algorithm, shown in Alg~\ref{algo:pd}, implements this idea in two phases.  In phase \RNum{1}, it collects in $\mb{C}$ the set of all pairs of
  variables that satisfy (a). Note that (a) is a symmetric property. At the end of this step $\mb{C}$ consists of all parents of $T$ and possibly parents of its children.
  In phase \RNum{2}, those  variables in $\mb{C}$ that violate (b) will be discarded,  in a single iteration over the subsets of $\mmb(T)$.  At the end of this step $\mb{C}$ consists of all and only parents of $T$.  While the \pdal\ algorithm uses principles similar 
to those used by the constrained-based CDD methods (Sec~\ref{sec:peri}), its local two-phase search strategy is novel
and  optimized for the discovery of  parents.  In contrast to other constraint-based algorithms that learn the structure of a DAG by first identifying
all direct neighbors of the nodes, the \pdal\ algorithm only checks whether a node is a neighbor of $T$ if it satisfies (a). In Sec.~\ref{sec:experi},  we  empirically show that our algorithm is more robust and efficient for covariates discovery than other CDD methods. The worst case complexity of the algorithm is exponential in the size of the largest Markov boundary it explores, which is typically much smaller than the number of attributes. \trv{In Ex.~\ref{ex:simp},  Markov Boundary of Carrier consists of only 5 out of 101 attributes in FlightData. Thus, our approach is effective for causal DAGs with bounded fan-ins. In Sec. \ref{sec:experi}, the largest Markov boundary computed in the experiments consists of 8 attributes.
}



\begin{algorithm}  
	\DontPrintSemicolon
	\KwIn{A dataset $D$,  and a treatment $T \in \att$}  
	\KwOut{A set of covariates $\mb{Z}$}
	$\mb{C} \gets \emptyset$\; 
	 \Comment{Phase \RNum{1}}\;
	\For{$Z \in \mmb(T) $ s.t. $Z \not \in \mb{C}$} {
			\For{$\mb{S} \subseteq \mmb(Z)-\{T\}$} { \label{li:par}
				\If{ $\exists \ W \in \mmb(T) $ s.t. $(Z \indep W \ | \mb{S}) \land (Z \nindep W \ | \ \mb{S} \cup \{T\}) $}{
					$ \mb{C} \gets  \mb{C}  \cup \{Z,W\}$\;
					\algorithmicbreak\;
				}
			}
	}
	 \Comment{Phase \RNum{2}}\;
	\For{$C \in \mb C$}{
	  		\If{ $\exists \ \mb S \subseteq \mmb(T)-\{C\} $ s.t. $(T \indep C \ | \mb{S})$}{
	  			$ \mb{C} \gets  \mb{C}  - \{C\}$\;
	  		}
  	}	
    $\mb Z \gets \mb C$\;
	\Return  $\mb{Z}$
	\caption{{\sc Covariate Detection (\pdal)}} 	\label{algo:pd}
\end{algorithm}

\ignore{
The $\pdal$ algorithm can be parameterized by restricting the search in Line \ref{li:par} to subsets of size $k$. It can be shown that if the number of
common children of the parents of $T$ are less than $k$, the parameterized version return the parents; of course if the assumption in Proposition \ref{prop:detec}
 are satisfied. We omit the details for lack of space.
}

\sys\ applies the \pdal\ algorithm to the subpopulation specifies by the WHERE clause of $\mc{Q}$, assuming homogeneity in the contexts formed by the grouping attributes. Note that for computing the direct effect of $T$ on each outcome $Y_j$, the parents of $\mb{PA}_{Y_j}$ must be also learned from data (Sec.~\ref{sec:peri}), using the \pdal\ algorithm.

{\bf Dropping logical dependencies.} As discuss in the introduction,
logical dependencies such as keys or functional dependencies can
completely confuse inference algorithms; for example, if the FD
$X \Rightarrow T$ holds then $\mmb(T) = \set{X}$, thus totally
isolating $T$ from the rest of the DAG.  \sys\ performs the following
steps.  Before computing the Markov boundary of a variable $T$ ,
it discards all attributes $X \in \att$ such that $H(T|X) = \epsilon$
and $H(X|T) = \epsilon$ for $\epsilon \approx 0$. These tests
correspond to approximate FDs, for example AirportWAC $\Rightarrow$
Airport. In addition, it drops attributes such as ID, FlightNum,
TailNum, etc., that have high entropy and either individually or in
combination form key constraints. Attributes with high entropy are either uninteresting or
must be further refined into finer categories.  One possibility is to choose a cut
point $\alpha$ and discard attributes $X \in \att$ with
$H(X) > \alpha$.  Instead, HypDB draws
	a set of small random samples from data, computes the entropy of
	all attributes in each sample, and uses an independence test to check
	whether entropies of the attributes in different samples depends
	on the sample sizes. The intuition is that entropy is a property of
	the underlying generative distribution Pr, not the size of a sample.
	For attributes such as ID, sample size functionally determines the
	entropy. Attributes with high entropy are either uninteresting or
	must be further refined into finer categories.




\begin{algorithm} 
	\DontPrintSemicolon
	\KwIn{ A dataset $D$, two variables $T,Y \in \att$ and a set $\mb Z \subset \att$,    number of permutation samples $m$ }
	\KwOut{ Significant level of $\hat{I}(T,Y|\mb{Z})$ }

	
	$s_0 \gets \hat{I}(X,Y|\mb{Z})$\;
	\For{$\mb{z} \in   \Pi_{\mb Z}(D)$} {
		\For{$i \in \{1 \ldots m \}$}{
		$CT_i\ \gets \mb{RandT}( \mc{C}^X_{\sigma_{\mb Z=\mb z}(D)}$, $\mc{C}^{Y}_{\sigma_{\mb Z=\mb z}(D)})$\;
		$S[\mb{z},i] \gets \hat{I}_{\mc{C}_i}(X,Y)$ \;
	}
	}
	$\hat{\alpha} \gets 0$\;	
	\For{$ i \in \{1, \dots, m\}$} {
		$s_i \gets 0$\;
		\For{$\mb{z} \in   \Pi_{\mb Z}(D))$} {  
			$s_i=+ S[\mb z,i]  \times \pr(\mb Z=\mb z)$
		}
		\If{$s_i >s_0$ }{$ \hat{\alpha}=+ \frac{1}{m} $.}
	}  
	$ \Return  \ \hat{\alpha} \mypm 1.96  \ \sqrt{\frac{\hat{\alpha}\ (1-\hat{\alpha})}{m}}$
	
		\caption{{\sc {\bf M}utual {\bf I}nformation {\bf T}est (\cmit)}} \label{algo:cmit}
\end{algorithm}

\section{Efficient Independence Test}
\label{sec:cit}

In this section we describe our algorithm for checking conditional
independence in the data.  The problem of determining significance of
dependency can be formulated as a chi-squared
test~\cite{greenwood1996guide}, G-test~\cite{mcdonald2009handbook} or
as exact tests such as the permutation
test~\cite{good2013permutation}. The permutation test applies to the
most general settings, and is non-parametric, but it is also
computationally very expensive.  In this section we propose new
optimization methods that significantly speedup the permutation test.
We start with the brief review.


\ignore{
parametric tests  are only valid in the asymptotic limit of infinite
data \cite{good2013permutation}.  \ignore{While the distribution-free nature of permutation tests makes them the most appropriate method for
hypothesis testing under a wide range of conditions,}  However, the  computational demands of exact test
can be runtime prohibitive. \ignore{,
especially if samples are not very small and/or many tests must be conducted.} This sections develops efficient permutation test  for testing conditional independence.
}

{\bf Monte-Carlo permutation test.}  $(T\indep Y |\mb{Z})$ holds iff
$I(T ; Y |\mb{Z})=0$, but in practice we can only estimate
$\hat{I}(T;Y| \mb{Z}) = v$. The aim of the permutation test is to
compute the p-value of a hypothesis test: under the null-hypothesis
$I(T;Y| \mb{Z})=0$, compute the probability that the estimate
$\hat{I}(T;Y| \mb{Z})$ is $\geq v$.  The distribution of the estimate
$\hat{I}$ under the null-hypothesis can be computed using the
following Monte-Carlo simulation: permute the values of the attribute
$T$ in the data within each group of the attributes $\mb Z$, re-compute
$\hat I$, and return the probability of $\hat I \geq v$.  In other
words, for each $\mb z \in \Pi_{\mb Z}(D)$, we compute
$\sigma_{\mb Z=\mb z}(D)$, randomly permute the values in the $T$
column (the permutation destroys any conditional dependence that may
have existed between $T$ and $Y$), then recompute $\hat I$, and set
the p-value $\alpha$ to the fraction of the $m$ trials where
$\hat I \geq v$.


The Monte Carlo simulation needs to be performed a sufficiently large
number of times, $m$, and each simulation requires permuting the
entire database.  This is infeasible even for small datasets.  Our
optimization uses contingency tables instead.


{\bf Permutation test using contingency tables.}  A contingency table
is a tabular summarization of categorical data. A k-way contingency
table over $\att$ is a $k$-dimensional array of non-negative integers
$\mc{C}^{\att}_D=\{n(\mb i)\}_{ \mb i \in \mb Dom(\att)}$, where
$n(\mb i)=\sum_{\satt \in D} 1_{\satt= \mb i}$. For any
$\mb{X} \subseteq \att$, marginal frequencies can be obtained by
summation over $\mb{X}$. For instance, the following table shows a
$2 \times 2$ contingency table over \ignore{two binary attributes} $T$
and $Y$ together with all marginal probabilities.

\begin{center} 

	\begin{tabular}{lcc|c}
		& $Y=1$ & $Y=0$ &  \\\hline
		$T=1$ & $n_{11}$ & $n_{10}$ & $n_{1\_}$\\ 
		$T=0$ & $n_{01}$ & $n_{00}$ & $n_{0\_}$\\\hline
		 & $n_{\_1}$ & $n_{\_0}$ & $n_{\_\_}$ \\ 
	\end{tabular}

\end{center}

\noindent

\ignore{
Monte-Carlo permutation samples are typically drawn by randomly shuffling the labels in data.  It is clear that random shuffling preserve the marginal frequencies but may change the joint frequencies. 
In fact, each random shuffling is amount to taking  a random sample from the distribution of  all possible contingency tables with fixed marginals.
}

  Randomly shuffling data only changes the entries of a contingency table, leaving  all marginal frequencies unchanged (or equivalently, shuffling data does not change the marginal entropies). Thus, instead of drawing random permutations
by shuffling data, one can draw them directly from the distribution of all contingency tables with fixed marginals. An efficient way to do this sampling is to use  Patefield's algorithm\cite{patefield1981algorithm}.
This algorithm accepts marginals of an $i \times j$ contingency table and generates
$m$ random contingency tables with the given marginals, where the probability of obtaining a table is the same as the probability of drawing it by randomly shuffling.

Based on this observation, we develop \cmit, a non-parametric test for significance of conditional mutual information, shown in Alg. \ref{algo:cmit}.  
To test the significance of $\hat{I}(T; Y|\mb Z)$,  for each $\mb z \in \Pi_{\mb Z}(D)$,   \cmit\ takes $m$ samples from the distribution of the contingency table
with fixed marginals $\mc{C}^X_{\sigma_{\mb Z=\mb z}(D)}$ and $\mc{C}^{Y}_{\sigma_{\mb Z=\mb z}(D)}$  using Patefield's algorithm.  Then, it  computes $\hat{I}_{C_i}(T;Y)$, the mutual information between
$T$ and $Y$ in the distribution defined by a random contingency table $C_i$. These results are aggregated using the equation $I(T; Y|\mb Z)= \E_{\mb{z}}[I(T; Y)| \mb{Z}=\mb{z}]$ to compute the test statistic in each permutation sample. 
Finally, \cmit\ computes a 95\% binomial proportion confidence interval around the observed p-value.

As opposed to the complexity of shuffling data, which is proportional to the size of the data, the complexity of  Patefield's algorithm
is proportional to the dimensions of $T$ and $Y$. Thus, the complexity of \cmit\ is essentially proportional to $m$, the number of permutation tests, and $|\Pi_{\mb Z}(D)|$, the number of groups.  
This makes \cmit\, orders of magnitude faster than the random shuffling of data. \trv{In Ex.~\ref{ex:simp} to test whether $\text{Carrier} \indep \text{Delayed}| \text{Airport}$, \cmit\ summarizes the data into four $2 \times 2$ contingency tables (one table per each airport), which is dramatically smaller than FlightData that consists of 50k rows.}


\ignore{
\begin{algorithm} \scriptsize
	\DontPrintSemicolon
	\KwIn{ A dataset $D$, and three subset of attributes $\bx$, $\mb{Y}$ and $\mb{Z}$, $k$ number of random permutation  sapmples}
	\KwOut{ Significant level of $I(\bx;\mb{Y}|\mb{Z})$ }
	\caption{{\sc Non-Parametric CMI test (\cmit)}}
	
	\label{algo:IAMB}
\end{algorithm}
}

{\bf Sampling from groups.}   If the dimension of the conditioning set $\mb Z$ becomes large, the curse of dimensionality  makes  \cmit\ infeasible. 
\ignore{Thus, these subsets are very small; hence, they make a little contribution to $I(X,T|\mb{Z})$. }
Let  $\hat{I}_{\mb z}$ be a random variables that represents the outcome of $\hat{I}_{C_i}(T;Y)$ for $\mb z \in \Pi_{\mb Z}(D)$. It holds that   $\hat{I}_{C_i}(T;Y) \leq \max(H(T|\mb Z=z_i),H(Y|\mb Z=z_i)$. Then, 
 the observed p-value $\alpha'$ reads as 
$P(a_0 \hat{I}_{\mb z_0}+ \ldots + a_c \hat{I}_{\mb z_c} \geq \hat{I}(T ; Y |\mb{Z}))$, where  $a_i= \pr(\mb Z=\mb z_i)$ and $c=|\Pi_{\mb Z}(D)|$. 
Thus,  a $ \mb z_i \in \Pi_{\mb Z}(D)$ with $w_i \defeq a_{z_i} \max(H(X|\mb Z=z_i),H(Y|\mb Z=z_i)) \approx 0$ does not affect the p-value.  Based on this observation, to further improve  performance, we restrict the test to a weighted sample of $\Pi_{\mb Z}(D)$, where the weights are $\{w_i\}$ for $i=1,c$.
 Note that for a fixed |$\Pi_{\mb Z}(D)$|, uniform random sampling is not effective. \cmit\ with sampling operates in an ``anytime" manner;  we empirically show that it is  reliable for small sampling fractions (Sec. \ref{sec:experi}). \ignore{In fact, sampling does not affect the false positive rate, but it may
  increase the change of false negative.} We leave the theoretical evaluation of
its precision for future work.

\vspace*{-0.2cm}
\section{Other Optimizations}
\label{sec:mbd}

We briefly report here other optimizations in \sys.


{\bf Materializing contingency tables.} The major computational efforts in all three components of \sys\ involve  contingency tables and computing entropies, which can be done by 
 $\texttt{count(*)} \ \texttt{Group By}$ query. However, this must be done for several 
 combinations of attributes. Contingency tables with their marginals are essentially OLAP data-cubes. 
Thus, with a pre-computed OLAP data cube,  \sys\ can detect, explain and resolve bias interactively at query time. In the absence of data-cubes, all three components of \sys\ can benefit from the on line materialization of selected contingency tables. For instance, in the \pdal\ (Alg.~\ref{algo:pd})
in both phases only the frequencies of a subset of attributes is required. In phase \RNum{1}, all  Independence tests are performed on a subset of $\mmb(Z) \cup \mmb(T)$; and In phase \RNum{2},
on a subset of $\mmb(Z) \cup \mb{C}$. Hence, \sys\ materializes appropriate contingency tables and compute the required marginal frequencies by summarization. Since contingency tables are materialized for attributes that are highly correlated, they are much smaller than the size of data.

{\bf Caching entropy.} A simple yet effective optimization employed by \sys\ is to cache entropies. Note that the 
computation of  $I(T;Y|Z)$ computes the entropies  $H(X)$, $H(Y)$, $ H(XZ)$ and $H(XYZ)$. These entropies are shared among many other conditional mutual information statements. For instance,
$H(T)$ and $ H(TZ)$  are shared between $I(T;Y|Z)$ and $I(T;W|Z)$. \sys\ caches entropies for efficient retrieval and to avoid redundant computations.

{\bf Hybrid independent test.} It is known that $\chi^2$ distribution
can be used for testing the significance of $\hat{I}(X;T|\mb Z)$, if
the sample size is sufficiently larger than the degree of freedom of
the test, calculated as
$df= (|\Pi_{X}(D)|-1)(|\Pi_{Y}(D)|-1) |\Pi_{\mb Z}(D)|$. Thus, \sys\,
uses the following hybrid approach for independent test: if
$ df \leq \frac{|D|}{\beta}$ ($\beta = 5$ is ideal) it uses the
$\chi^2$ approximation; otherwise, it performs permutation test using
\cmit.  We call this approach \hcmit.


\section{Experimental Results}
\label{sec:experi}
We implemented \sys\, in Python to use it
as a standalone library.
This section presents experiments that evaluate the feasibility and
efficacy of \sys.  We aim to address the following questions.
{\bf Q1}:  To what extent  \sys\ does prevent the chance of false discoveries?
{\bf Q2}: What are the end-to-end results of \sys?
{\bf Q3}:  What is the quality of the automatic covariate discovery
algorithm in \sys, and how does it compare to the state of the art CDD methods?  {\bf Q4:} What is the efficacy of the proposed optimization techniques?

\begin{table*}[] \centering 
	\begin{tabular}{@{}lrrrrrrrrr@{}}\toprule
		{Dataset} & {Columns [$\#$]} & {Rows[$\#$]} &  Det. & Exp. & Res.    \\ \midrule
		\textbf{AdultData}~\cite{adult} & 15 & 48842 & 65  & <1 & <1  \\ \hdashline
		\textbf{StaplesData}~\cite{valentino2012websites}& 6  & 988871 & 5 & <1 & <1\\ \hdashline
		\textbf{BerkeleyData}~\cite{bickel1975sex} & 3 & 4428  &  2 & <1 & <1  \\ \hdashline
		\textbf{CancerData}~\cite{LUCAS} & 12 & 2000 & <1 & <1 & <1  \\ \hdashline
		
		\textbf{FlightData} ~\cite{flightdata} & 101 & 43853 & 20 & <1 & <1 \\ \midrule
		
	\end{tabular}
	\caption{Runtime in seconds for detection, explanation and resolution of bias in  experiments in Sec.~\ref{sec:endtoend}.}
	\footnotetext{ Note that the list reports only a portion of FlightData relevant to Ex.~\ref{ex:simp}. Average number of attributes categories (distinct values) are computed for attributes that are not doped by \sys\ due to high entropy}
	\label{tbl:data}
\end{table*}

\vspace*{-0.2cm}
\subsection{Setup}
\label{sec:setup}



{\bf Data.}  For  $\bf (Q1)$ we used 50M entries in the FlightData.
Table~\ref{tbl:data} shows the  datasets we used for $\bf (Q
2)$.  For  $\bf (Q3)$ and $\bf (Q
4)$, we needed ground truth for quality comparisons, so we generated more than 100 categorical datasets of varying sizes for which the underlying causal DAG is known.
To this end, we  first generated a set of random DAGs using the Erd\H{o}s-R\`enyi model.  The DAGs were generated with 8, 16 and 32 nodes, and the expected number of edges was in the range 3-5. Then, each DAG was seen
as a causal model that encodes a set of conditional independences.  Next, we drew samples from the distribution defined by these DAGs using the catnet package in R~\cite{balov2016use}. Note that causal DAGs admit the same factorized distribution as Bayesian networks \cite{pearl2009causal}.
The samples were generated with different sizes in the range 10K-501M rows, and different numbers of attribute categories (numbers of distinct values) were in the range 2-20.
We refer to these datasets as  RandomData. 

\ignore{
	The {\em FlightData} we used was collected by the US
	Department of Transportation (U.S. DOT) \cite{flightdata}. It contains
	records of more than 90\% of US domestic flights of major airlines
	from 1988 to the present. \ignore{Table \ref{tab:attlist}(shown in Section
		\ref{sec:introduction}) lists dataset attributes that are relevant to
		our experiments. } We restrict our analysis to about 10M data entry
	collected between 2000 and 2015.  The {\em BerkeleyData}, consists of three attributes Gender, Accepted and Department and
	contains graduate admissions decisions at
	University of Berkeley  for 4,425 applicants \cite{bickel1975sex}.  The {\em AdultData}
	which  reports census data and income  for
	48,842 U.S. citizens.
}

{\bf Significance test.}  We used \cmit\ with 1000 permutations to test the significance of the differences between the answers to the queries $\mc{Q}$ (Listings~\ref{olap}) and $\rwq$ (Listings~\ref{rfq}). It is easy to see that
the difference is zero iff $I(T;Y)=0$ for $\mc{Q}$ and iff $I(Y;T|\mb{Z})=0$ for $\rwq$.

{\it Systems.} The experiments were performed locally on a 64-bit OS X machine with an
Intel Corei7 processor (16 GB RAM, 2.8 GHz).


\ignore{
	\begin{figure}[t!]
		\vspace*{-1cm} \hspace*{-1cm}  \includegraphics[scale=1.1]{fig/fdr.pdf}
		\caption{\bf{End-to-End Performance of \sys.}}\label{fig:graph}
	\end{figure}
}

\vspace*{-0.2cm}
\subsection{Avoiding false discoveries ({\bf Q1})} \label{sec:fd}
\begin{figure*}[t!]
	\begin{subfigure}{0.49\textwidth}
		\includegraphics[height=4cm,width=1\linewidth]{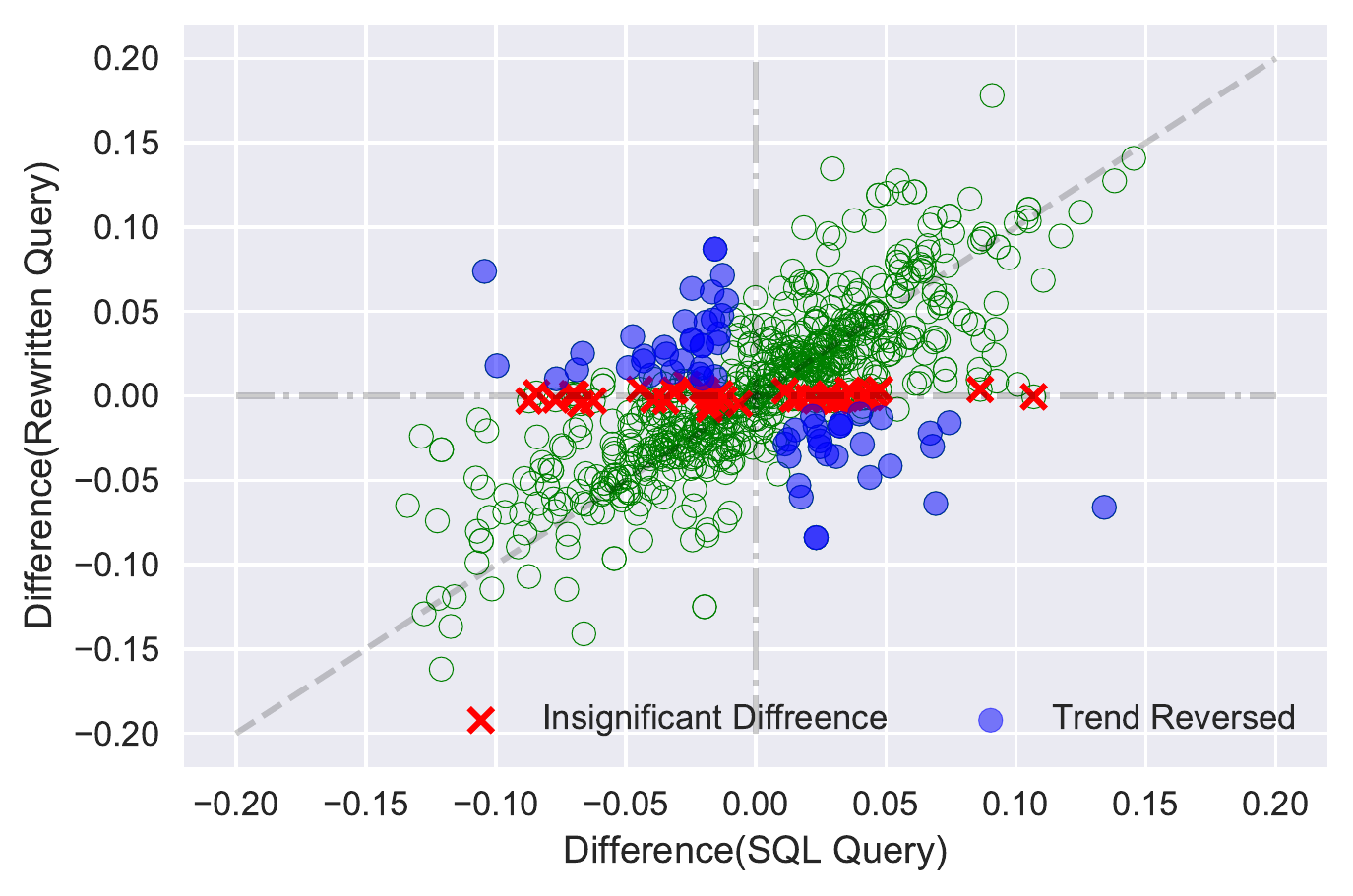}
		\caption{The effect of query rewriting on FlightData.}
	\end{subfigure}
	\begin{subfigure}{0.49\textwidth}
		\includegraphics[height=4cm,width=1\linewidth]{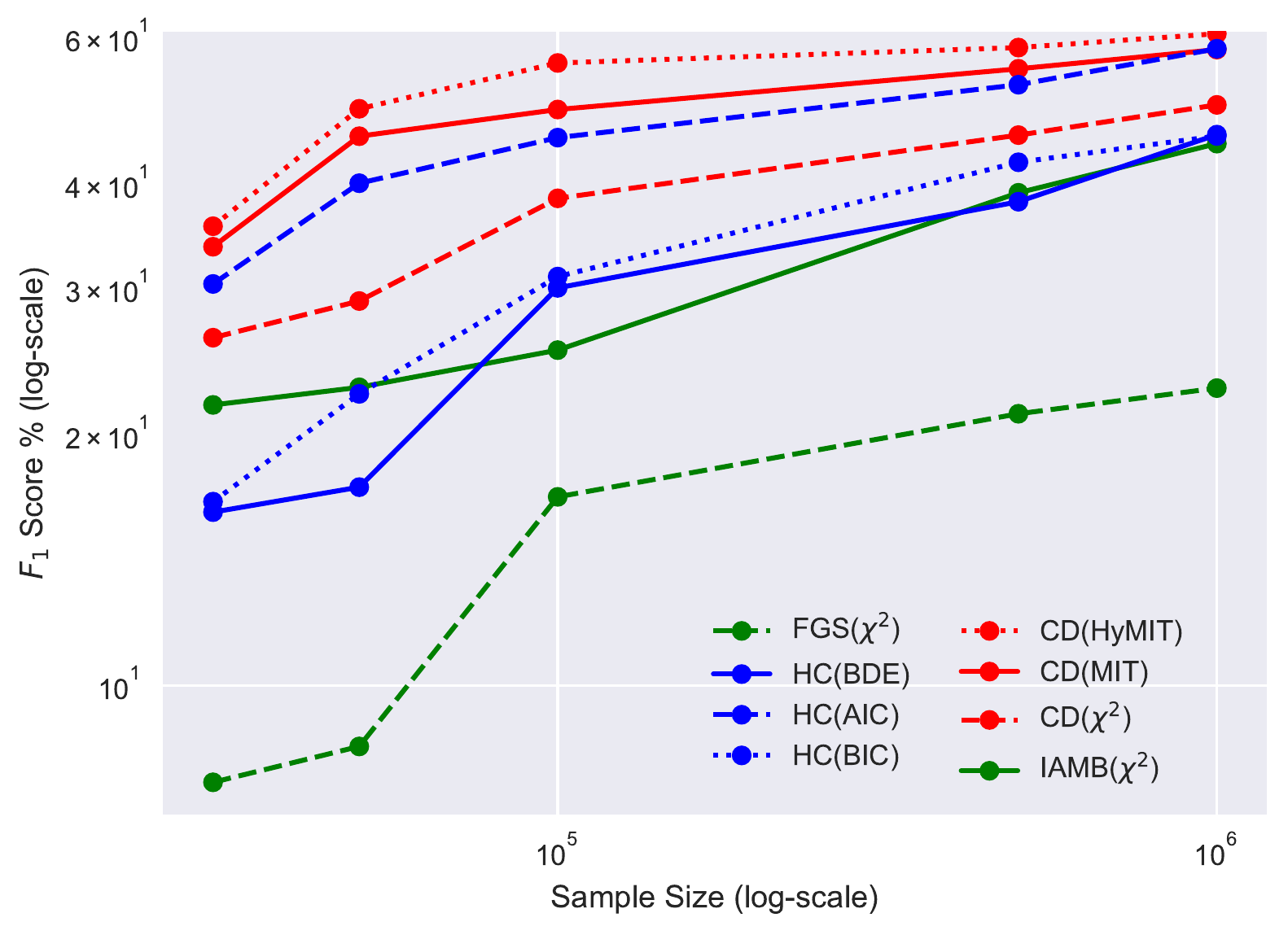}
		\caption{Quality comparison.}
	\end{subfigure}

	\begin{subfigure}{0.49\textwidth}
		\vspace*{1mm}
		\includegraphics[height=4cm,width=1\linewidth]{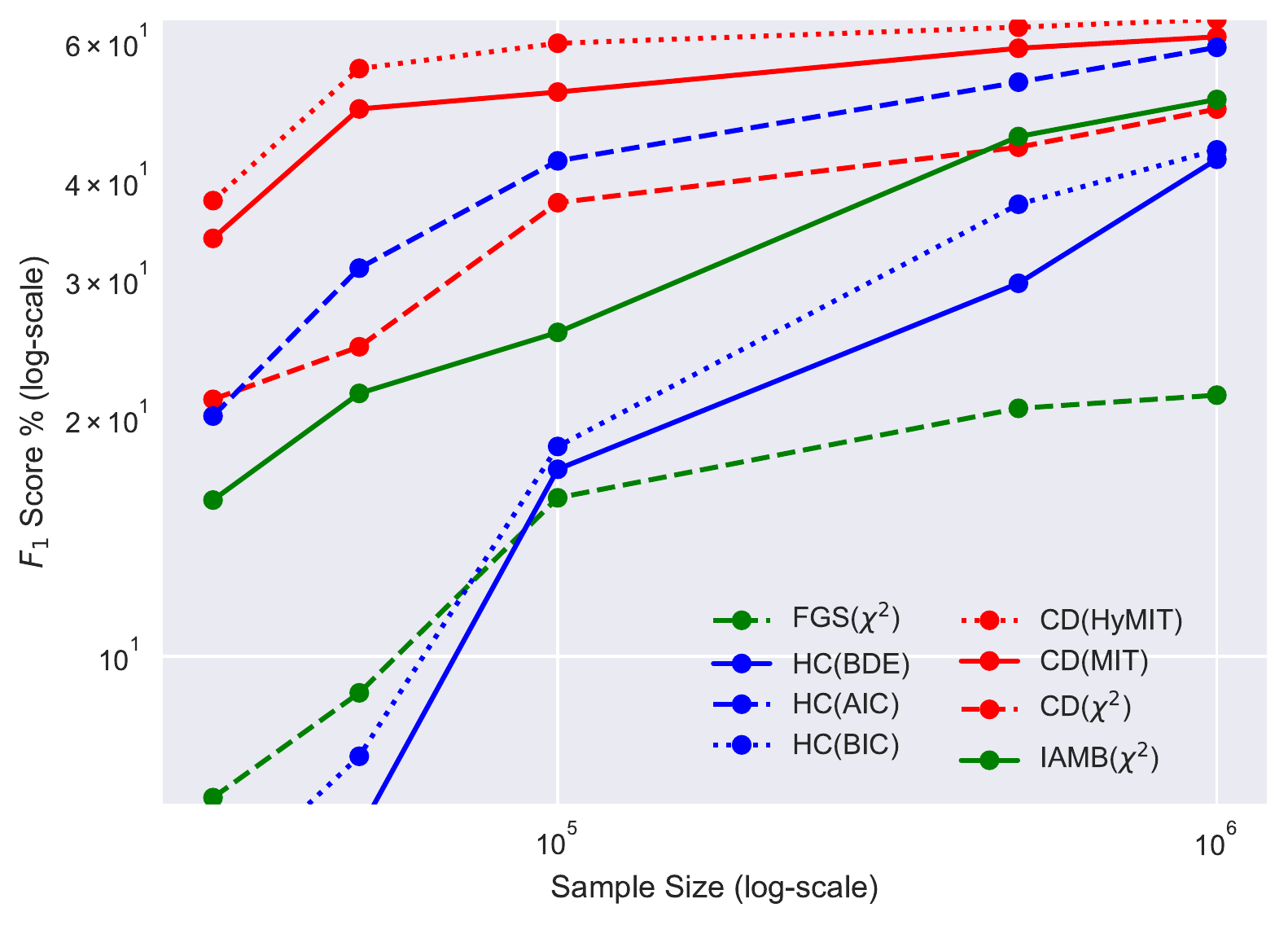}
		\caption{Quality comparison for nodes with at least two parents.}
	\end{subfigure}
	\begin{subfigure}{0.49\textwidth}
		\includegraphics[height=4cm,width=1\linewidth]{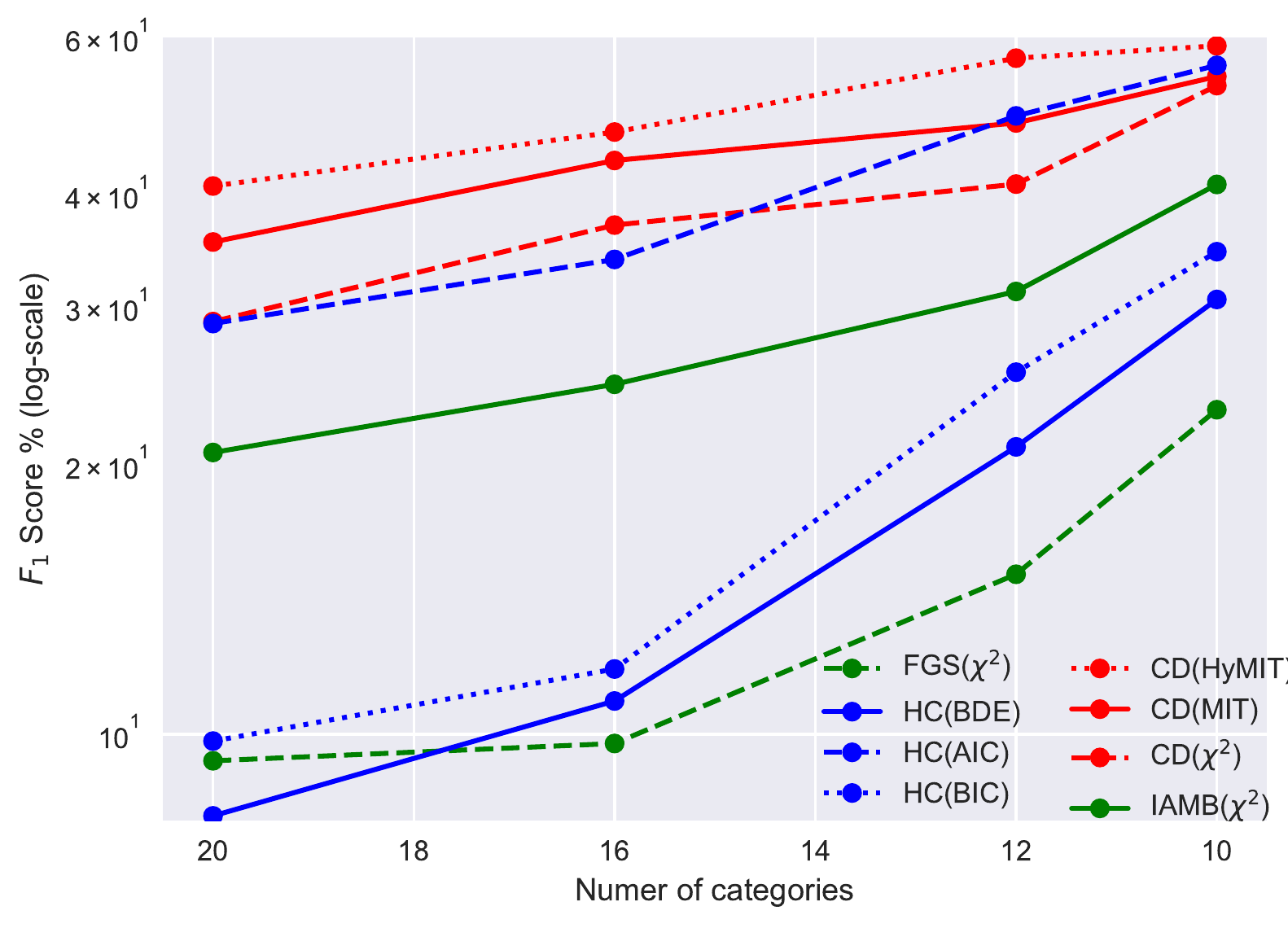}
		\caption{Quality comparison with decreasing number of categories.}
	\end{subfigure}
	\caption{\bf Results of the experiments.}
	\label{fig:graphs1}
\end{figure*}

What are the chances that a data analyst does a false discovery by running a SQL query?
For this experiment, we generated 1000 random SQL queries of the form $\mc{Q}$ (Listing~\ref{olap}) from FlightData (queries with random, months, airports, carriers, etc.) that compare the performance of two carriers (similar to Ex.~\ref{ex:simp}).
We used \sys\ to rewrite the queries into a query of the form $\rwq$ w.r.t. the potential covariates Airport, Day,  Month, DayOfWeek.
\frv{As shown in  Fig~\ref{fig:graphs1} (a), for more than 10\% of SQL queries that indicate a significant difference between the performance of carriers, the difference became insignificant after query rewriting. That is, the observed difference in such cases explained by the covariates. Fig~\ref{fig:graphs1} (a) also shows
in 20\% of the cases, query rewriting reversed the trend (similar to Ex. \ref{ex:simp}). Indeed, for any query that is not located in the diagonal of the graph in Fig~\ref{fig:graphs1} (a), query rewriting was effective.}

\ignore{In any dataset, unless the distribution of the chosen $T$ is not affected by any variables in the domain, $\mc{Q}$ is biased and leads to false discoveries.
	The experiments  in the next section confirm this. }
\vspace*{-0.2cm}

\begin{figure}[t!] \centering 
	\includegraphics[scale=0.6]{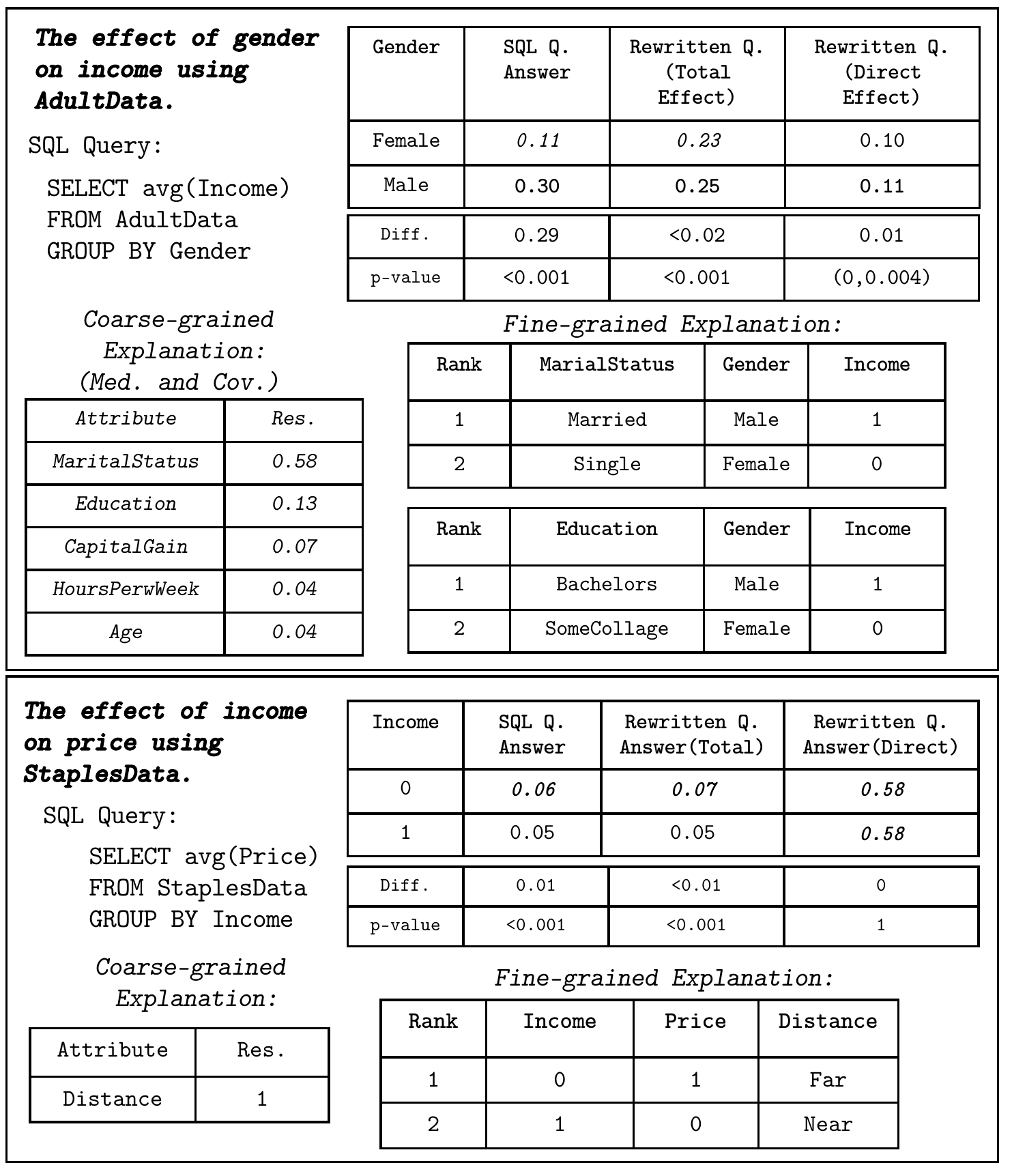}
	\caption{\bf The effect of gender  on income in AdultData (top); The effect of income on price in StaplesData (bottom). }
	\label{fig:deprep2}
\end{figure}[t!]

\subsection{End-to-end results {\bf (Q2)}} \label{sec:endtoend}  In the following experiments, for each query, the relevant covariates and mediators were detected using the \pdal\ algorithm with \hcmit\ (Sec.~\ref{sec:mbd}).
Table.~\ref{tbl:data} reports the running times of the covariates detection.  \frv{ Some of the datasets  used in this experiment
also investigated by FairTest \cite{tramer2017fairtest}. By
using the same datasets, we could  compare our results and confirm them.}

{\bf AdultData.} Using this dataset, several prior works in algorithmic fairness have reported  gender discrimination
based on a strong statistical dependency between income and gender in favor of males \cite{luong2011k,vzliobaite2011handling,tramer2017fairtest}. \ignore{In legal dispute, however, the target of litigation
	is whether sex has any direct effect of income \cite{pearl2001direct}.} \frv{In particular, FairTest reports 11\% of women have
	high income compared to 30\% of men, which suggests a huge disparity against women.} We applied \sys\ to AdultData to compute the  effect of gender on income. We started with the query in Fig. \ref{fig:deprep2} (top), which computes the average of Income (1 iff Income> 50k) Group By Gender, which indeed suggests  a strong
disparity with respect to females' income. \frv{Note that FairTest essentially reports the result of the query in Fig.~\ref{fig:deprep2} (top).
In contrast,  \sys\  detects that this query is biased. It identifies attributes, such as  MaritalStatus,  Education, Occupation and etc., as mediators and  covariates.  The result of the rewritten query
suggests that the disparity between male and female is not nearly as drastic. The explanations generated by \sys\ show that Maritalstatus accounts for most of the bias, followed by Education.  However, the top fine-grained explanations for MaritalStatus reveal  surprising facts: there are more married males in the data than married females, and marriage has a strong positive association with high income.
It turns out that the income attribute in US census data reports the adjusted gross income as indicated in the individual's tax forms, which depends on  filing status (jointly and separately), could be  household income. Thus,
{\em AdultData is inconsistent and should not be used to investigate gender discrimination.}} \sys\  explanations also show that males tend  to have higher educations than females and higher educations is associated with higher incomes. \frv{Although, this dataset does not meet the assumptions needed for inferring causal conclusions, \sys's  report is illuminating and goes beyond FairTest.}

\begin{figure}[t!] \centering
	
	\includegraphics[scale=0.6]{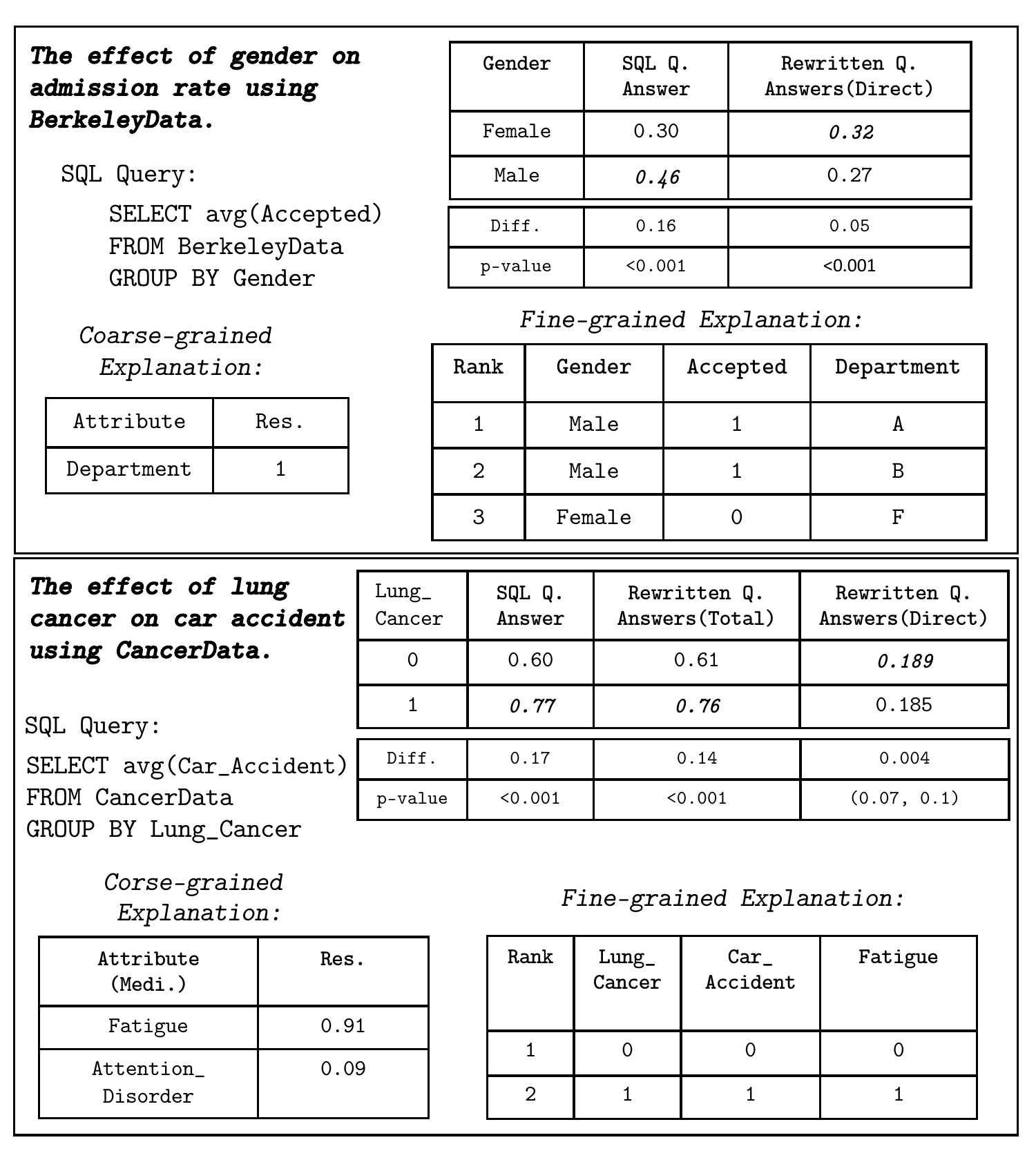}
	\caption{\bf  Report of the effect of lung cancer on car accident on CancerData (top);  report of the effect  of gender discrimination in BerkeleyData (bottom). }
	\label{fig:deprep1}
\end{figure}

{\bf BerkeleyData.}  In 1973, UC Berkeley was sued for discrimination
against females in graduate school admissions. The admission figures
for the fall of 1973 showed that men applying were more likely than
women to be admitted, and the difference was so large that it was
unlikely to be due to chance. The result of the query in
Fig.~\ref{fig:deprep1} (top) suggests a huge disparate impact on
female applicants.  However, the query is bias
w.r.t. Department. After removing bias by rewriting the query, \sys\
reveals that disparity between males and females is not nearly as
drastic (Fig. \ref{fig:deprep1} (top)).  Note that potentially missing
covariates, such as an applicant's qualification, prohibits causal
interpretation of the answers. However, the fine-grained explanations
generated by \sys\ are insightful. They reveal that females tended to
apply to departments such as F that have lower acceptance rates,
whereas males tended to apply to departments such as A and B that have
higher acceptance rates. \ignore{, in which the authors concluded:
  {\em women tended to apply to departments with low rates of
    admission, whereas men tended to apply to less-competitive
    departments with high rates of admission.}}
\frv{For BerkeleyData, FairTest reports a strong association between
  Gender and Acceptance, which becomes insignificant after
  conditioning on Department. In contrast, \sys\ reveals that there
  still exists an association even after conditioning, but the trend
  is reversed! In addition, \sys's explanations demystify the seemingly
  paradoxical behavior of this dataset. These explanations agree with
  the conclusion of \cite{bickel1975sex}, in which the authors
  investigated BerkeleyData.  }

{\bf StaplesData.}
{\em Wall Street Journal} investigators showed
that Staples' online pricing algorithm discriminated against
lower income people \cite{valentino2012websites}. The situation was referred to as an
``unintended consequence"  of Staples's seemingly rational
decision to adjust online prices based on user proximity to
competitors' stores.  We used \sys\ to investigate the problem. As depicted in Fig~\ref{fig:deprep2} (bottom),  \sys\ reveals that Income has no direct effect
on Price. However, it has an indirect effect via Distance. The explanations show that this is simply because
people with low incomes tend to live far from  competitors' stores, and
people who live far get higher prices. This is essentially the conclusion of \cite{valentino2012websites}.  \frv{ For StaplesData, FairTest reports strong association
	between Income and Price, which is confirmed by \sys. However, the obtained insights from \sys, e.g., the indirect interaction of Income and Price, are more profound and critically important for answering the question whether the observed discrimination is intended or unintended.}

{\bf CancerData.}  This is a simulated dataset
generated according to the causal DAG shown in
Fig.~\ref{fig:exdag} in the appendix.  This data was used to test all three components of  \sys\ against ground truth. We used the query in Fig.~\ref{fig:deprep1} (bottom) to decide whether lung cancer has any impact on car accidents.
According to the ground truth, there is no direct edge between lung cancer and car accidents; hence, there is no significant direct causal effect. However, since there is an indirect path between them, we expect a significant total causal effect.
As shown in Fig.~\ref{fig:deprep1} (bottom),  \sys\ detects that this query is biased and correctly discovers sufficient confounding and mediator variables.
The explanations for bias show that fatigue is the most responsible attribute for bias; people with lung cancer tend to be fatigued, which is highly associated with car accidents.  Thus, the answers to the rewritten queries and explanations coincide with the ground truth.

{\bf FlightData.}  The results presented in Ex.~\ref{ex:simp} were generated using \sys. During covariate detection, \sys\ identifies and drops logical dependencies
induced by attributes such as  FlightNum, TailNum, AirportWAC,  etc.  It identified
attributes such as Airport, Year, ArrDelay, Deptime, etc., as covariates and mediating variables. The generated explanations coincide with the intuition in Ex.~\ref{ex:simp}.

Finally, we remark that the assumptions needed to perform parametric independence tests fail for  FlightData and AdultData due to the large number of categories in their attributes and the high density of the underlying causal DAG; (this also justifies the higher running time for these datasets). The analysis of these datasets was possible only with the non-parametric tests developed in Sec~\ref{sec:cit}. (Also, other CDD methods we discuss in the next section were not able to infer sufficient covariates and mediators.) To test the significance of $\hat{I}(T;Y|
\mb Z)$ with \cmit, the permutation confidence interval was computed based on $m=100$ permutation samples. We restricted the test to a sample of groups of size proportional to $ \log(|\Pi_{\mb Z}(D)|)$
as described in Sec.~\ref{sec:cit}. Note that we used the significance level of 0.01 in all statistical tests in the
\pdal\ algorithm.

\vspace*{-0.2cm}

\subsection{Quality comparison ({\bf Q3})}
\label{sec:qc}
We used RandomData, for which we had the ground truth, to  assess the quality of the \pdal\ algorithm. We used the algorithm to learn the parents of all attributes in the corresponding DAG underlying different datasets in RandomData.
We repeated the experiment with the following independence tests: \cmit\ with sampling (same sampling fraction as in Sec~\ref{sec:endtoend}), \hcmit\ and $\chi^2$.
We used the F1-score as the accuracy metric to measure the performance the \pdal\ algorithm and compared it to the following reference algorithms implemented in the bnlearn library in R \cite{nagarajan2013bayesian}: two contained-based methods, Full Grow-Shrink (FGS) \cite{margaritis2000bayesian} and Incremental Association (IAMB) \cite{tsamardinos2003algorithms} with $\chi^2$ independent test; the score based greedy search
with Akaike Information Criterion (AIC), Bayesian Dirichlet equivalent (BDe) and  Bayesian Information Criterion (BIC) scores.  The significance level of 0.01 was used in all statistical tests. \ignore{Note that the experiment is restricted to datasets of size 1M, because reference algorithms can not handle larger datasets.}

The FGS utilizes Markov boundary for learning the structure of a causal DAG. It first discovers the Markov boundary  of all nodes using the Grow-Shrink algorithm. Then, it determines the underlying undirected graph, which consists of all nodes and their neighbors. For edge orientation, it uses similar principles
as used in the \pdal\ algorithm. The IAMB
is similar to FGS except that it uses an improved version of the Grow-Shrink algorithm to learn Markov boundaries. Note that the superiority of CDD methods based on Markov boundary to other constraint-based method (such as the  PC algorithm \cite{spirtes2000causation}) was shown in \cite{pellet2008using}. Thus, we restricted the comparison to these algorithms.

Fig. \ref{fig:graphs1} (b) shows that our algorithm significantly outperforms most other algorithms. We remark, however, that this comparison is not fair, because the  \pdal\ algorithm is not designed for learning the entire structure of a DAG.  In fact, other constrained-based algorithms use the information across different nodes for edge origination. Thus, they could potentially learn the parent of a nodes with only one parent. Since learning the entire DAG is not the focus
of our algorithm, in Fig. \ref{fig:graphs1} (c) and (d) we restrict the comparison to nodes with at least two parents (either neighbors or not). As depicted, the \pdal\ algorithm with \hcmit\ outperforms all other algorithms. Notice that for smaller datasets and larger number of categories, our algorithm performs much better than the other algorithms. In fact, for a fixed DAG,  $\chi^2$ test and score based methods become less reliable on sparse data.  \frv{ Conditioning on $\mb Z$ splits the data into $|\Pi_{\mb Z}(D)|$ groups. Thus, conditioning on large $\mb Z$ causes the data to split into very small groups that makes inference about independence less reliable.  Fig. \ref{fig:graphs1} (d) shows that , for sparse data,  tests based on permutation deliver highest accuracy.  Note
	that size conditioning sets in the \pdal\ algorithm depends on the density of underling causal DAGs.
	In Ex.~\ref{ex:simp},  the largest conditioning set used by \sys,  consists of only 6 out of 101 attributes in FlightData.}



\begin{figure*}[t!]
	\begin{subfigure}{0.49\textwidth} \centering
		\hspace*{0.22cm}\vspace*{.2cm}
		\includegraphics[height=4.cm,width=1\linewidth]{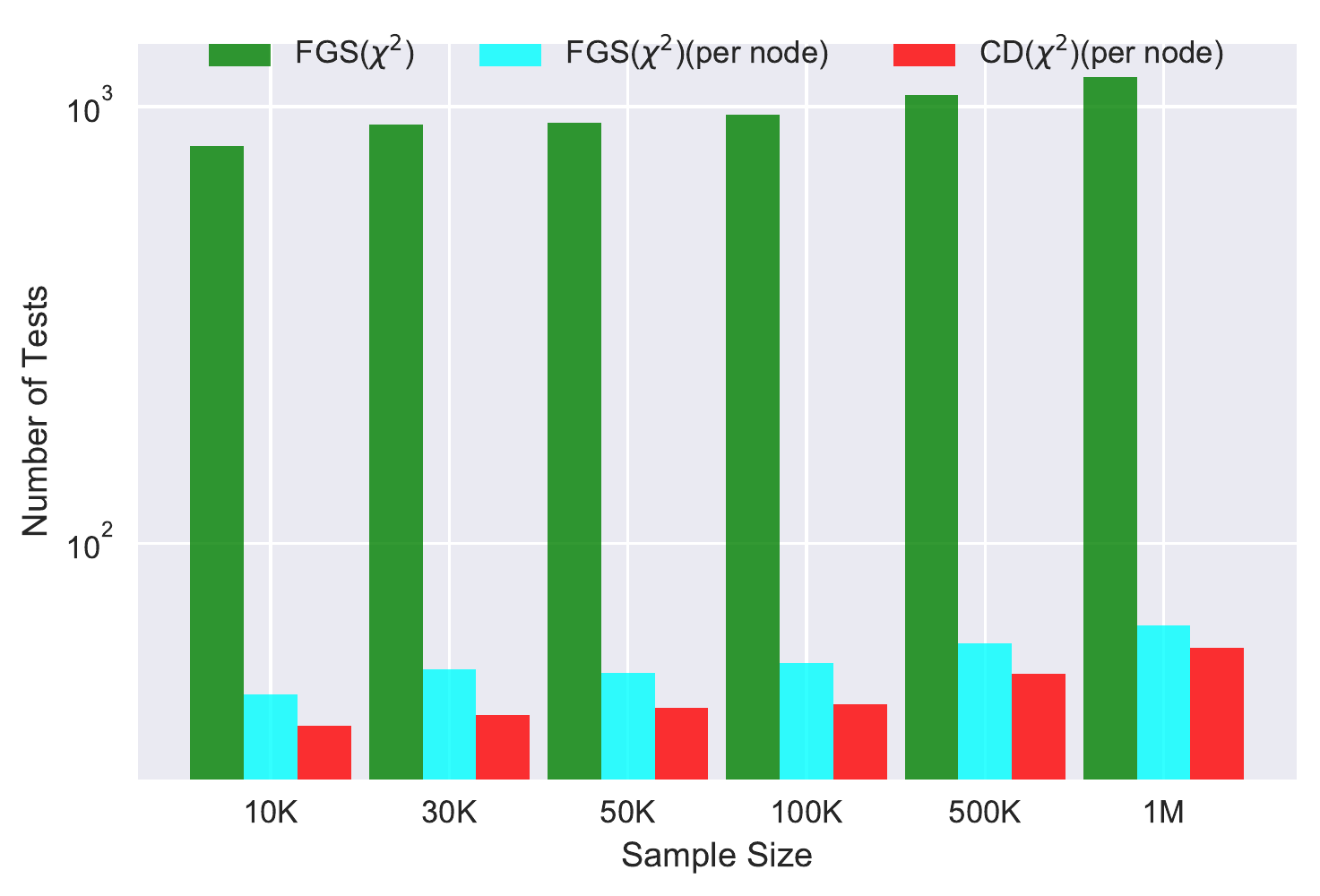}
		\vspace*{-0.1cm}\caption{Comparing the number of independence tests.}
	\end{subfigure}
	\begin{subfigure}{0.49\textwidth}
		\includegraphics[height=4cm,width=1\linewidth]{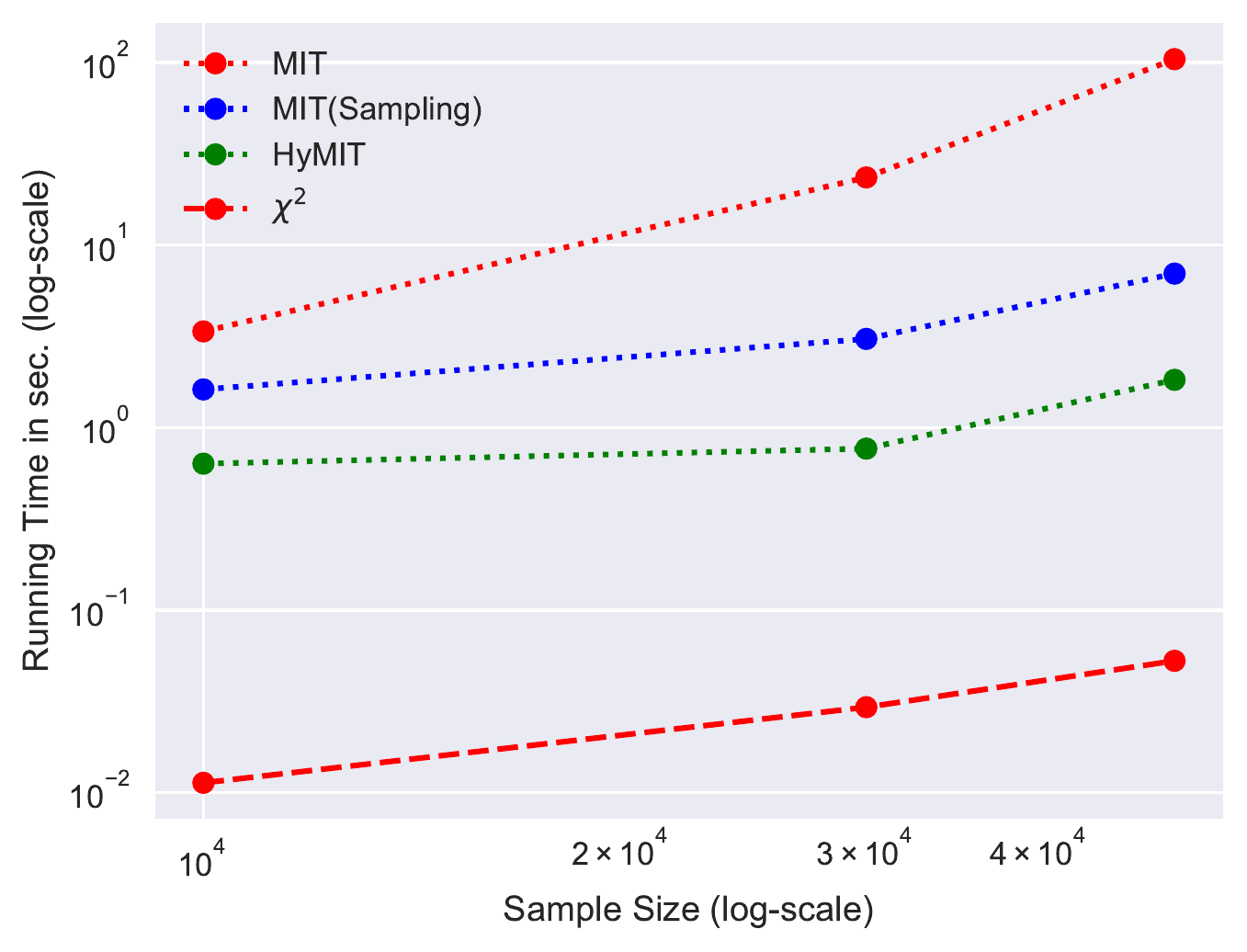}
		\caption{Efficacy of the optimizations proposed for independence tests.}
	\end{subfigure}
	\begin{subfigure}{0.49\textwidth}
	\hspace{0.0cm}	\includegraphics[height=4cm,width=1\linewidth]{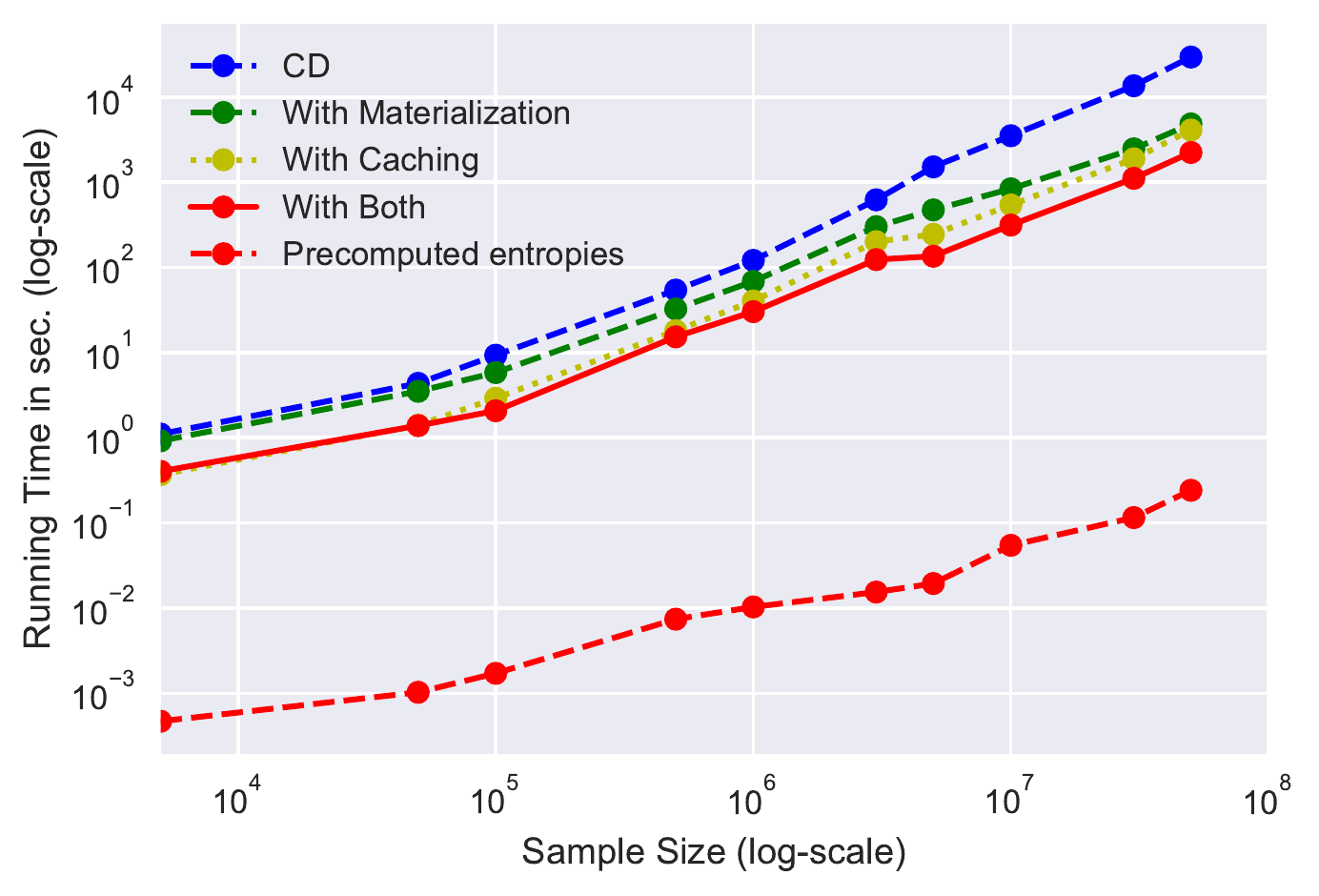}
		\caption{Efficacy caching and materialization.}
	\end{subfigure}
	\begin{subfigure}{0.49\textwidth}
	\includegraphics[height=4cm,width=1\linewidth]{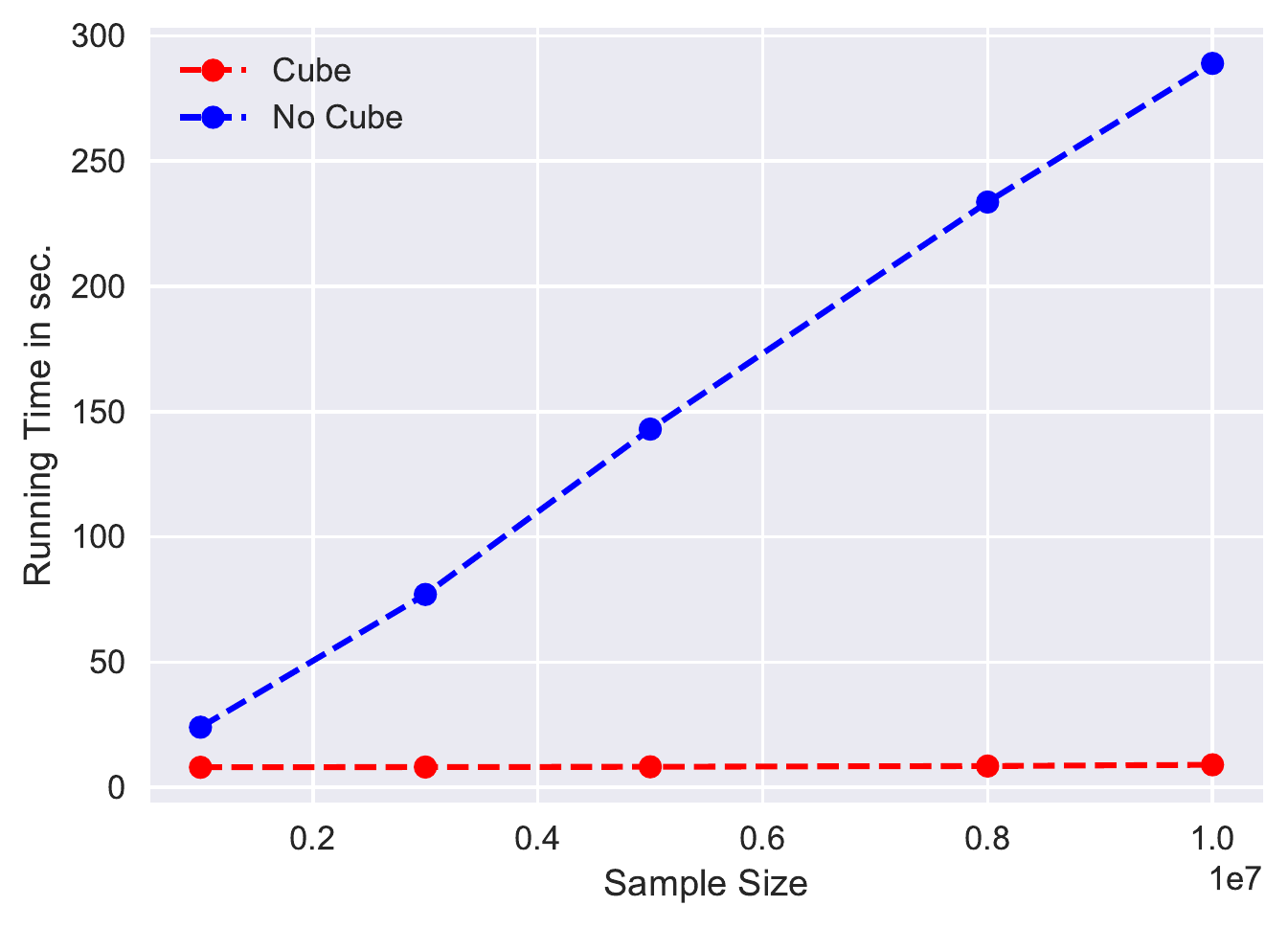}
	\caption{Benefits of using data cubes.}
\end{subfigure}
	\caption{\bf Results of the experiments.}
	\label{fig:graphs2}
\end{figure*}

An interesting observation is that even though our method uses
principles that are similar to the other constraint-based methods, it outperforms them even with the same independence test, i.e., $\chi_2$ test. This is because
the \pdal\  algorithm uses a novel two-phase search strategy that optimized for learning parents, and does not relay on learning the underling undirected graph.   As shown in Fig~\ref{fig:graphs2} (a),
the \pdal\ algorithm conducted fewer independence tests per node than the FGS algorithm.  Fewer independence tests
not only improve efficiency but  make the algorithm more reliable.\footnote{ The number of  conducted  independence tests is typically reported as a measure of the performance of CDD methods.}
Note that we only compared with the FGS, because we also used the Grow-Shrink algorithm to compute Markov boundaries.
Also, notice that  learning the parents of a nodes required many fewer independence test than the entire causal DAG. This makes our algorithm scalable to highly dimensional data for which learning the entire causal DAG is infeasible.

\ignore{
	We are convinced of the superiority of the Markov blanket approaches as described in this paper.
	We invoke as support for this claim the high run times of PC, and the good low and high sample size
	accuracy of GS and TC/TCbw, respectively. Not only are Markov blanket techniques much more
	scalable, they can be more accurate; they are also more easily modifiable to construct only parts the
	network deemed relevant by some criterion.
	http://www.jmlr.org/papers/volume9/pellet08a/pellet08a.pdf
}

\vspace*{-0.1cm}

\subsection{Efficacy of the optimization techniques ({\bf Q4})}
\label{sec:opti_expri}
To evaluate the quality of the optimizations proposed for non-parametric independence tests, we compared the running time and performance of \cmit, \cmit\ with sampling  (same sampling fraction as in Sec~\ref{sec:endtoend}),  \hcmit\ and $\chi^2$ tests using the same data as used in Sec~\ref{sec:qc},
but we restricted the experiments to samples smaller than 50k, to study their behavior on sparse data. Fig~\ref{fig:graphs2} (b) compares  the average running time of performing each tests. As depicted, both \cmit\ with sampling and  \hcmit\ are much faster that \cmit.   Fig~\ref{fig:addplots} (a), in the appendix, shows that the proposed tests have comparable accuracy. Note that \hcmit\ performs better
than  \cmit\ with sampling, because it avoids sampling when $\chi^2$ test is applicable.
 Performing one permutation test with shuffling data consumes hours in the smallest dataset used in this experiment,
whereas with \cmit\ takes less than a second.

To evaluate the efficacy of materializing contingency tables and caching entropies, we used the same data as in Sec~\ref{sec:qc}. As shown in Fig.~ \ref{fig:graphs2} (c), both optimizations are effective.
The efficacy of materializing
contingency tables increases for larger sample sizes, because these tables becomes relatively much smaller than the size of data as the sample size increases. Note that we used pandas \cite{mckinney2011pandas} to implement the statistical testing framework of \sys. Computing entropies, which is essentially a group-by query,  with pandas is up to 100 times slower
than the same task in bnlearn.  Fig.~ \ref{fig:graphs2} (c) also shows the running time of the \pdal\ algorithm minus the time spent for computing entropies. As depicted, entropy computation constitutes the major computational effort in the  \pdal\
algorithm.

\srv{
Finally, we showed that computation of \pdal\ algorithm can benefit from pre-computed OLAP data-cubes and can be pushed inside
a database engine.  We used PostgreSQL to pre-compute data-cubes (with Count as measure)
for RandomData with 8, 10 and 12
attributes.   In Fig.~ \ref{fig:graphs2} (d),  we vary the input data size,
whereas  in Fig~\ref{fig:addplots} (b), in the appendix, we vary the number of
attributes. Both the graphs show that
the advantage of using data-cube is dramatic. Note that the cube operator in PostgreSQL is restricted to 12 attributes.
We also restrict to binary data for which computing a cube on the largest dataset used in this experiment took up to 30 hours. Without restricting to binary we could only compute a
cube over a few number of attributes and small datasets.}

\ignore{
 Thus, with a pre-computed OLAP data cube,  \sys can detect bias
interactively at query time.}

\section{Discussion and Related Work}
\label{sec:diss}

\paragraph*{\bf Assumptions.}  \sys\ can detect, explain and resolve bias of a query under three assumptions: (1) parents of the treatment attributes
	in the underling causal DAG are included in data, (2) the treatment has at least two non-neighbor parents, and (3) The faithfulness assumption (see Sec. \ref{sec:app_ab}), which implies conditional independence implies no casual relationship. While drawing causal conclusions without (1) is impossible in general, there are techniques to handle unobserved attributes in certain cases \cite{pearl2009causality}, that can be incorporated in \sys. Failure of (2) fundamentally prohibits the identification  of the parents from observational data.
Sec.~\ref{sec:acd} offers an agenda for future work to deal with such cases. The failure of (3) has been the subject of many (philosophical) discussions. However, it has been argued that in most practical settings this assumption is justifiable (see \cite{neufeld2005whether}).

\paragraph*{\bf Algorithmic fairness.} While it is known in causal inference that any claim of unfairness requires evidence of causality \cite{pearl2001direct}, most work in algorithmic fairness defines discrimination as strong statistical dependency between an algorithm's outputs and protected attributes. For instance in legal dispute over hiring discrimination, neither the correlation between sex or race and hiring, nor their effect
 on applicant's qualification, nor the effect of qualifications on hiring are target of investigation. Rather, to prove discrimination one must show that race or sex directly affect hiring decisions. Even though they may indirectly affect hiring by way of  applicant qualification. In several experiments, we showed that one can use \sys\ to detect unfairness post factum using simple SQL queries.  
\sys\ reach for beyond state-of-the-art tools for fairness such as Fairtest \cite{tramer2017fairtest}.




\paragraph*{\bf Statistical Errors.} \frv{\sys\ relies on conditional independent tests that are subject to false positives and false negatives. While it is generally impossible to absolutely prevent the two types of errors simultaneously, there are standard techniques to control for the false discovery rate in learning causal DAGs  (e.g., \cite{li2009controlling}). We leave this extension for future work. \ignore{Note that we empirically showed that the non-parametric independence test developed in this paper
are more robust than the parametric test for sparse data (see Sec.~\ref{sec:qc}).} The ramification of statistical errors in interpreting \sys\ results can be summarized as follows: if \sys\ reports no significant effect after rewriting a SQL query wrt. inferred covariates, then its  answers are reliable under Faithfulness and if sufficient data is available, regardless of statistical errors and potential spurious attributes in the covarites. The reason is that the set of covariates at hand explains the spurious correlation reported by the SQL query.  However, if the obtained effect after query rewriting is still significant, then in the presence of statistical errors, the answers could be unreliable.}

\paragraph*{\bf OLAP Aspects.} More general OLAP queries e.g., queries with drill-down, roll-up or cube operator can be expressed as a set of group-by queries as studied in this paper. However, addressing the performance issues goes beyond the scope of this paper. In the context of causality, 
measures such as average treatment effect, likelihood ratios, analysis of variance (ANOVA) and information-theoretic metrics (such as mutual information) often used to quantify causal effect (see,~\cite{janzing2013quantifying}). \sys\ can be extended to support these aggregates. Finally,  the type of analysis proposed is this paper is not supported by OLAP data cubes. However, we showed that 
pre-computed cubes significantly speed up all components of \sys. They rely on computing conditional probabilities or entropies that are GROUP BY count queries.

\paragraph*{\bf Simpson's Paradox.}
Prior research \cite{fabris2006discovering,freitas2001understanding,glymour1997statistical,guo2017you},
studied Simpson's paradox in OLAP and data mining. They concerned with the efficient detection of instances of Simpson's paradox as unexpected/surprising patterns in data. However, it is widely 
known in causal inference that such statistical anomalies neither reveal interesting facts in data nor are paradoxical \cite{pearl2011simpson}.  They appear paradoxical once the result of biased queries is given causal interpretation and occur as a result of ignoring confounding variables. \ignore{In contrast,  \sys\ avoids statistical anomalies pre factum by automatically inferring confounding variables and  removing bias from queries. }

\paragraph*{\bf Hypothetical Queries.}
Much literature addresses hypothetical OLAP queries,  e.g., \cite{balmin2000hypothetical,lakshmanan2008if, deutch2013caravan}. They concerned with computing OLAP queries given a hypothetical scenario, which updates a database. 
 Their approach, however 
is not adequate for data-driven decision making. Databases are the observed outcome of complicated stochastic processes in which the causal mechanism 
that determines the value of one variable interferes with those that determine others; hence, computing the effect of  hypothetical updates requires knowledge about the causal interaction of variables. Thus, \sys\ learns parts of the causal model relevant to a query at hand to account for confounding influences. Our future work includes extending \sys\ to efficiently answer arbitrary ``what-if"
and ``how-so" queries that drive actionable insights.

\paragraph*{\bf Causality in databases}
The notion of causality has been studied extensively in databases \cite{MeliouGMS2011,roy2014formal,SalimiTaPP16,DBLP:conf/icdt/SalimiB15,bertossi2017causes2,bertossi2017causes,salimi2015query}.  We note that this line of work is different than the present paper in the sense that, it aims to identify causes for an observed output of a data transformation. For example, in query-answer causality/explanation, the objective is to identify parts of a database that explain a result of a query. While these works share some aspects of the notion of causality as studied in this paper, the problems that they address are fundamentally different. In \cite{salimi2017zaliql} it has been shown that common inference problems 
in causality can be pushed into database system.

\section{Conclusion}
This paper proposed \sys, a system to detect, explain, and
resolve bias in decision-support OLAP queries.  We showed that biased queries can be perplexing and lead to statistical anomalies, such as Simpson's paradox.  We proposed a
novel technique to find explanations for the bias, thereby assisting the
analyst in interpreting the results.  We developed an automated
method for rewriting the query into an unbiased query that correctly
performs the hypothesis test that the analyst had in mind.  The rewritten queries compute
causal effects or the effect of hypothetical interventions. At the core of our framework lies the ability to find confounding variables. We showed that our method outperforms the state of the art causal DAG discovery methods.  We showed that  \sys\ can be used to 
detect algorithmic unfairness post factum and  the
obtained insights go beyond state of the art e.g., Fairtest
\cite{tramer2017fairtest}. Our system can be used
as an adhoc analysis along with OLAP data-cubes to detect, resolve and explain bias interactively at query time.

\bibliographystyle{plain}
\bibliography{ref}

\begin{thebibliography}{10}

\bibitem{balmin2000hypothetical}
Andrey Balmin, Thanos Papadimitriou, and Yannis Papakonstantinou.
\newblock Hypothetical queries in an olap environment.
\newblock In {\em VLDB}, volume 220, page 231, 2000.

\bibitem{balov2016use}
Nikolay Balov and Peter Salzman.
\newblock How to use the catnet package, 2016.

\bibitem{bertossi2017causes}
Leopoldo Bertossi and Babak Salimi.
\newblock Causes for query answers from databases: datalog abduction,
  view-updates, and integrity constraints.
\newblock {\em International Journal of Approximate Reasoning}, 90:226--252,
  2017.

\bibitem{bertossi2017causes2}
Leopoldo Bertossi and Babak Salimi.
\newblock From causes for database queries to repairs and model-based diagnosis
  and back.
\newblock {\em Theory of Computing Systems}, 61(1):191--232, 2017.

\bibitem{bickel1975sex}
Peter~J Bickel, Eugene~A Hammel, J~William O’Connell, et~al.
\newblock Sex bias in graduate admissions: Data from berkeley.
\newblock {\em Science}, 187(4175):398--404, 1975.

\bibitem{Binnig2017TowardSI}
Carsten Binnig, Lorenzo~De Stefani, Tim Kraska, Eli Upfal, Emanuel Zgraggen,
  and Zheguang Zhao.
\newblock Toward sustainable insights, or why polygamy is bad for you.
\newblock In {\em CIDR}, 2017.

\bibitem{de2011covariate}
Xavier De~Luna, Ingeborg Waernbaum, and Thomas~S Richardson.
\newblock Covariate selection for the nonparametric estimation of an average
  treatment effect.
\newblock {\em Biometrika}, page asr041, 2011.

\bibitem{deutch2013caravan}
Daniel Deutch, Zachary~G Ives, Tova Milo, and Val Tannen.
\newblock Caravan: Provisioning for what-if analysis.
\newblock In {\em CIDR}, 2013.

\bibitem{fabris2006discovering}
Carem~C Fabris and Alex~A Freitas.
\newblock Discovering surprising instances of simpson's paradox in hierarchical
  multidimensional data.
\newblock {\em International Journal of Data Warehousing and Mining (IJDWM)},
  2(1):27--49, 2006.

\bibitem{foster2004causation}
Sheila~R Foster.
\newblock Causation in antidiscrimination law: Beyond intent versus impact.
\newblock {\em Hous. L. Rev.}, 41:1469, 2004.

\bibitem{freitas2001understanding}
Alex~A Freitas.
\newblock Understanding the crucial role of attribute interaction in data
  mining.
\newblock {\em Artificial Intelligence Review}, 16(3):177--199, 2001.

\bibitem{freitas2006we}
Alex~A Freitas.
\newblock Are we really discovering interesting knowledge from data.
\newblock {\em Expert Update (the BCS-SGAI magazine)}, 9(1):41--47, 2006.

\bibitem{glymour1997statistical}
Clark Glymour, David Madigan, Daryl Pregibon, and Padhraic Smyth.
\newblock Statistical themes and lessons for data mining.
\newblock {\em Data mining and knowledge discovery}, 1(1):11--28, 1997.

\bibitem{good2013permutation}
Phillip Good.
\newblock {\em Permutation tests: a practical guide to resampling methods for
  testing hypotheses}.
\newblock Springer Science \& Business Media, 2013.

\bibitem{greenwood1996guide}
Priscilla~E Greenwood and Michael~S Nikulin.
\newblock {\em A guide to chi-squared testing}, volume 280.
\newblock John Wiley \& Sons, 1996.

\bibitem{guo2017you}
Yue Guo, Carsten Binnig, and Tim Kraska.
\newblock What you see is not what you get!: Detecting simpson's paradoxes
  during data exploration.
\newblock In {\em Proceedings of the 2nd Workshop on Human-In-the-Loop Data
  Analytics}, page~2. ACM, 2017.

\bibitem{LUCAS}
Isabelle Guyon.
\newblock Lung cancer simple model, 10 2009.

\bibitem{heckerman1998tutorial}
David Heckerman et~al.
\newblock A tutorial on learning with bayesian networks.
\newblock {\em Nato Asi Series D Behavioural And Social Sciences}, 89:301--354,
  1998.

\bibitem{holland1986statistics}
Paul~W Holland.
\newblock Statistics and causal inference.
\newblock {\em Journal of the American statistical Association},
  81(396):945--960, 1986.

\bibitem{Holland1986}
Paul~W. Holland.
\newblock Statistics and causal inference.
\newblock {\em Journal of the American Statistical Association}, 81(396):pp.
  945--960, 1986.

\bibitem{iacus2009cem}
Stefano~M Iacus, Gary King, Giuseppe Porro, et~al.
\newblock Cem: software for coarsened exact matching.
\newblock {\em Journal of Statistical Software}, 30(9):1--27, 2009.

\bibitem{janzing2013quantifying}
Dominik Janzing, David Balduzzi, Moritz Grosse-Wentrup, and Bernhard
  Sch{\"o}lkopf.
\newblock Quantifying causal influences.
\newblock {\em The Annals of Statistics}, 41(5):2324--2358, 2013.

\bibitem{lakshmanan2008if}
Laks~VS Lakshmanan, Alex Russakovsky, and Vaishnavi Sashikanth.
\newblock What-if olap queries with changing dimensions.
\newblock In {\em Data Engineering, 2008. ICDE 2008. IEEE 24th International
  Conference on}, pages 1334--1336. IEEE, 2008.

\bibitem{li2009controlling}
Junning Li and Z~Jane Wang.
\newblock Controlling the false discovery rate of the association/causality
  structure learned with the pc algorithm.
\newblock {\em Journal of Machine Learning Research}, 10(Feb):475--514, 2009.

\bibitem{adult}
M.~Lichman.
\newblock Uci machine learning repository, 2013.

\bibitem{lin2010rank}
Shili Lin.
\newblock Rank aggregation methods.
\newblock {\em Wiley Interdisciplinary Reviews: Computational Statistics},
  2(5):555--570, 2010.

\bibitem{luong2011k}
Binh~Thanh Luong, Salvatore Ruggieri, and Franco Turini.
\newblock k-nn as an implementation of situation testing for discrimination
  discovery and prevention.
\newblock In {\em Proceedings of the 17th ACM SIGKDD international conference
  on Knowledge discovery and data mining}, pages 502--510. ACM, 2011.

\bibitem{margaritis2000bayesian}
Dimitris Margaritis and Sebastian Thrun.
\newblock Bayesian network induction via local neighborhoods.
\newblock In {\em Advances in neural information processing systems}, pages
  505--511, 2000.

\bibitem{mcdonald2009handbook}
John~H McDonald.
\newblock {\em Handbook of biological statistics}, volume~2.
\newblock Sparky House Publishing, 2009.

\bibitem{mckinney2011pandas}
Wes McKinney.
\newblock pandas: a foundational python library for data analysis and
  statistics.
\newblock {\em Python for High Performance and Scientific Computing}, pages
  1--9, 2011.

\bibitem{MeliouGMS2011}
Alexandra Meliou, Wolfgang Gatterbauer, Katherine~F. Moore, and Dan Suciu.
\newblock The complexity of causality and responsibility for query answers and
  non-answers.
\newblock {\em {Proc. VLDB Endow. (PVLDB)}}, 4(1):34--45, 2010.

\bibitem{miller1955note}
George~A Miller.
\newblock Note on the bias of information estimates.
\newblock {\em Information theory in psychology: Problems and methods},
  2(95):100, 1955.

\bibitem{nagarajan2013bayesian}
Radhakrishnan Nagarajan, Marco Scutari, and Sophie L{\`e}bre.
\newblock Bayesian networks in r.
\newblock {\em Springer}, 122:125--127, 2013.

\bibitem{neapolitan2004learning}
Richard~E Neapolitan et~al.
\newblock {\em Learning bayesian networks}, volume~38.
\newblock Pearson Prentice Hall Upper Saddle River, NJ, 2004.

\bibitem{neufeld2005whether}
Eric Neufeld and Sonje Kristtorn.
\newblock Whether non-correlation implies non-causation.
\newblock In {\em FLAIRS Conference}, pages 772--777, 2005.

\bibitem{patefield1981algorithm}
WM~Patefield.
\newblock Algorithm as 159: an efficient method of generating random r$\times$
  c tables with given row and column totals.
\newblock {\em Journal of the Royal Statistical Society. Series C (Applied
  Statistics)}, 30(1):91--97, 1981.

\bibitem{pearl1993bayesian}
Judea Pearl.
\newblock [bayesian analysis in expert systems]: Comment: Graphical models,
  causality and intervention.
\newblock {\em Statistical Science}, 8(3):266--269, 1993.

\bibitem{pearl2001direct}
Judea Pearl.
\newblock Direct and indirect effects.
\newblock In {\em Proceedings of the seventeenth conference on uncertainty in
  artificial intelligence}, pages 411--420. Morgan Kaufmann Publishers Inc.,
  2001.

\bibitem{pearl2009causality}
Judea Pearl.
\newblock {\em Causality}.
\newblock Cambridge university press, 2009.

\bibitem{pearl2010introduction}
Judea Pearl.
\newblock An introduction to causal inference.
\newblock {\em The international journal of biostatistics}, 6(2), 2010.

\bibitem{pearl2011simpson}
Judea Pearl.
\newblock Simpson's paradox: An anatomy.
\newblock {\em Department of Statistics, UCLA}, 2011.

\bibitem{pearl2014probabilistic}
Judea Pearl.
\newblock {\em Probabilistic reasoning in intelligent systems: networks of
  plausible inference}.
\newblock Morgan Kaufmann, 2014.

\bibitem{pearl2009causal}
Judea Pearl et~al.
\newblock Causal inference in statistics: An overview.
\newblock {\em Statistics Surveys}, 3:96--146, 2009.

\bibitem{pedreshi2008discrimination}
Dino Pedreshi, Salvatore Ruggieri, and Franco Turini.
\newblock Discrimination-aware data mining.
\newblock In {\em Proceedings of the 14th ACM SIGKDD international conference
  on Knowledge discovery and data mining}, pages 560--568. ACM, 2008.

\bibitem{pellet2008using}
Jean-Philippe Pellet and Andr{\'e} Elisseeff.
\newblock Using markov blankets for causal structure learning.
\newblock {\em Journal of Machine Learning Research}, 9(Jul):1295--1342, 2008.

\bibitem{flightdata}
Dataset: Airline On-Time Performance.
\newblock {{\url{http://www.transtats.bts.gov/}}}.

\bibitem{Spirtes:book01}
Clark~Glymour Peter~Spirtes and Richard Scheines.
\newblock {\em Causation, Prediction and Search}.
\newblock MIT Press, 2001.

\bibitem{roy2014formal}
Sudeepa Roy and Dan Suciu.
\newblock A formal approach to finding explanations for database queries.
\newblock In {\em Proceedings of the 2014 ACM SIGMOD international conference
  on Management of data}, pages 1579--1590. ACM, 2014.

\bibitem{rubin1970thesis}
Donald~B Rubin.
\newblock {\em The Use of Matched Sampling and Regression Adjustment in
  Observational Studies}.
\newblock Ph.D. Thesis, Department of Statistics, Harvard University,
  Cambridge, MA, 1970.

\bibitem{rubin1986statistics}
Donald~B Rubin.
\newblock Statistics and causal inference: Comment: Which ifs have causal
  answers.
\newblock {\em Journal of the American Statistical Association},
  81(396):961--962, 1986.

\bibitem{salimi2015query}
Babak Salimi.
\newblock {\em Query-Answer Causality in Databases and its Connections with
  Reverse Reasoning Tasks in Data and Knowledge Management}.
\newblock PhD thesis, Carleton University, 2015.

\bibitem{SalimiTaPP16}
Babak Salimi, Leopoldo Bertossi, Dan Suciu, and Guy {Van den Broeck}.
\newblock Quantifying causal effects on query answering in databases.
\newblock In {\em TaPP}, 2016.

\bibitem{DBLP:conf/icdt/SalimiB15}
Babak Salimi and Leopoldo~E. Bertossi.
\newblock From causes for database queries to repairs and model-based diagnosis
  and back.
\newblock In {\em ICDT}, pages 342--362, 2015.

\bibitem{salimi2017zaliql}
Babak Salimi, Corey Cole, Dan~RK Ports, and Dan Suciu.
\newblock Zaliql: causal inference from observational data at scale.
\newblock {\em Proceedings of the VLDB Endowment}, 10(12):1957--1960, 2017.

\bibitem{salimi2018bias}
Babak Salimi, Johannes Gehrke, and Dan Suciu.
\newblock Bias in olap queries: Detection, explanation, and removal.
\newblock In {\em Proceedings of the 2018 International Conference on
  Management of Data}, pages 1021--1035. ACM, 2018.

\bibitem{spirtes2000causation}
Peter Spirtes, Clark~N Glymour, and Richard Scheines.
\newblock {\em Causation, prediction, and search}.
\newblock MIT press, 2000.

\bibitem{tramer2017fairtest}
Florian Tramer, Vaggelis Atlidakis, Roxana Geambasu, Daniel Hsu, Jean-Pierre
  Hubaux, Mathias Humbert, Ari Juels, and Huang Lin.
\newblock Fairtest: Discovering unwarranted associations in data-driven
  applications.
\newblock In {\em Security and Privacy (EuroS\&P), 2017 IEEE European Symposium
  on}, pages 401--416. IEEE, 2017.

\bibitem{tsamardinos2003algorithms}
Ioannis Tsamardinos, Constantin~F Aliferis, Alexander~R Statnikov, and
  Er~Statnikov.
\newblock Algorithms for large scale markov blanket discovery.
\newblock In {\em FLAIRS conference}, volume~2, pages 376--380, 2003.

\bibitem{valentino2012websites}
Jennifer Valentino-Devries, Jeremy Singer-Vine, and Ashkan Soltani.
\newblock Websites vary prices, deals based on users’ information.
\newblock {\em Wall Street Journal}, 10:60--68, 2012.

\bibitem{zemel2013learning}
Rich Zemel, Yu~Wu, Kevin Swersky, Toni Pitassi, and Cynthia Dwork.
\newblock Learning fair representations.
\newblock In {\em Proceedings of the 30th International Conference on Machine
  Learning (ICML-13)}, pages 325--333, 2013.

\bibitem{vzliobaite2011handling}
Indre {\v{Z}}liobaite, Faisal Kamiran, and Toon Calders.
\newblock Handling conditional discrimination.
\newblock In {\em Data Mining (ICDM), 2011 IEEE 11th International Conference
  on}, pages 992--1001. IEEE, 2011.

\end{thebibliography}


\newpage
\section{Appendix}

\subsection{Additional Background}
\label{sec:app_ab}

We give here some more technical details to the material in Sec.~\ref{sec:peri}.

\paragraph*{\bf Entropy} The {\em entropy} of a subset of random
variables $\bx \subseteq \att$ is
$H(\bx) \defeq -\sum_{\mb x \in Dom(\mb X)} \pr(\mb x) \log \pr(\mb
x)$,
where $\pr(\mb x)$ is the marginal probability. The {\em conditional
  mutual information (CMI)} is
$I(\bx;\mb{Y}|\mb{Z})\defeq
H(\bx\mb{Z})+H(\bx\mb{Y})-H(\bx\mb{Z}\mb{Y})+H(\mb{Z})$.
We say that $\bx$, $\mb{Y}$ are {\em conditionally independent (CI)},
in notation $\bx \indep \mb{Y}| \mb{Z}$, if $I(\bx;\mb{Y}|\mb{Z})=0$.
Notice that all these quantities are defined in terms of the unknown
population and the unknown probability $\pr(\att)$.  To estimate the
entropy from the database $D$ we use the Miller-Madow estimator
\cite{miller1955note}:
$\hat{H}(\bx)= \sum_{ \sx \in \Pi_{\bx}(D)} F(\sx) \log F(\sx) +
\frac{m-1}{2n} $,
where $F(\sx)= \frac{1}{n} \sum_{\satt \in D} 1_{\satt[\bx]=\sx}$ (the
empirical distribution function) and $m=|\Pi_\bx(D)|$ is the number of
distinct elements of $\bx$. We refer to the sample estimate of
$I(\bx;\mb{Y}|\mb{Z})$ as $\hat{I}(\bx;\mb{Y}|\mb{Z})$.

\paragraph*{\bf Justification of Unconfoundedness} In the Neyman-Rubin
Causal Model, the {\em independence assumption} states that
$(Y(t_0),Y(t_1) \bigCI T)$.  This assumption immediately implies
$\E[Y(t_i)] = \E[Y(t_i)|T=t_i]$, $i=0,1$, and therefore $\ate$ can be
computed as:
\begin{align}
  \ate(T,Y) = \E[Y|T=t_1] - \E[Y|T=t_0] \label{eq:adj0}
\end{align}
Here $Y$ is $Y(T)$, the attribute present in the data, and thus
Eq.~\eqref{eq:adj} can be estimated from $D$ as the difference of
$\texttt{avg}(Y)$ for $T=t_1$ and for $T=t_0$.  Notice that one should
not to confuse the independence assumption
$({Y(t_0), Y(t_1) \bigCI T})$ with $({Y \bigCI T})$, meaning
$({Y(T) \bigCI T})$; if the latter holds, then $T$ has no causal
effect on $Y$.  Under the independence assumption,

The independence assumption holds in {\em randomized} data (where the
treatment $T$ is chosen randomly), but fails in {\em observational}
data.  In the case of observational data, we need to rely on the
weaker Assumption~\ref{assumption:1}.  Notice that Unconfoundedness
essentially states that the independence assumption holds for each
value of the covariates.  Thus, Eq.(\ref{eq:adj0}) holds once we
condition on the covariates, proving the adjustment formula
(\ref{eq:adj}).

\paragraph*{\bf Causal DAGs} Intuitively, a {\em causal DAG} $G$ with
nodes $V(G)=\att$, and edges $E(G)$ captures all potential causes
between the
variables~\cite{pearl2001direct,pearl2009causality,de2011covariate}.
We review here how compute the covariates $\mb Z$ using the DAG $G$,
following~\cite{pearl2009causal}.  A node $X_i$ is a {\em parent} of
$X_j$ if $(X_i,X_j) \in E(G)$, $\mb{PA}_{X_j}$ denotes the set of
parents of $X_j$, and two nodes $X_i$ and $X_j$ are {\em neighbors} if
one of them is a parent of the other one. A {\em path} $\mb{P}$ is a
sequence of nodes $X_1, \ldots, X_\ell$ such that $X_i$ and $X_{i+1}$
are neighbors forall $i$.  $\mb P$ is {\em directed} if
$(X_i,X_{i+1}) \in E(G)$ forall $i$, otherwise it is {\em
  nondirected}.  If there is a directed path from $X$ to $Y$ then  we write
$X \stackrel{*}{\rightarrow} Y$, and we
say $X$ is an {\em ancestor}, or a {\em cause} of $Y$, and $Y$ is a
{\em descendant} or an {\em effect} of $X$.  A nondirected path
$\mb P = (X_1, \ldots, X_\ell)$ from $X_1$ to $X_\ell$ is called a
{\em back-door} if $(X_2, X_1)\in E(G)$ and
$(X_{\ell -1}, X_\ell) \in E(G)$.  $X_k$ is a {\em collider} in a path
$\mb P$ if both $X_{k-1}$ and $X_{k+1}$ are parents of $X_k$.  A path
with a collider is {\em closed}; otherwise it is {\em open}; note that
an open path has the form
$X \stackrel{*}{\leftarrow} \stackrel{*}{\rightarrow} Y$, i.e.  $X$
causes $Y$ or $Y$ causes $X$ or they have a common cause. If $\mb P$
is open, then we say that a set of nodes $\mb Z$ {\em closes} $\mb P$
if ${\mb P} \cap {\mb Z} \neq \emptyset$. 
Given two sets of nodes $\mb{X},\mb{Y}$ we say that a set $\mb{Z}$
{\em d-separates}\footnote{d stands for ``directional''.}  $\mb X$ and
$\mb Y$, denoted by $\mb{X} \indep \mb{Y} |_d \ \mb{Z}$, if $\mb Z$
closes every open path from $\mb{X}$ to
$\mb{Y}$~\cite{pearl2009causal}.  \trv{This special handling of
  colliders, reflects a general phenomenon known as Berkson's paradox,
  whereby conditioning on a common consequence of two independent
  cause render spurious correlation between them, see
  Ex. \ref{ex:coll} below.}

\begin{figure*}  \center
	\includegraphics[scale=0.35]{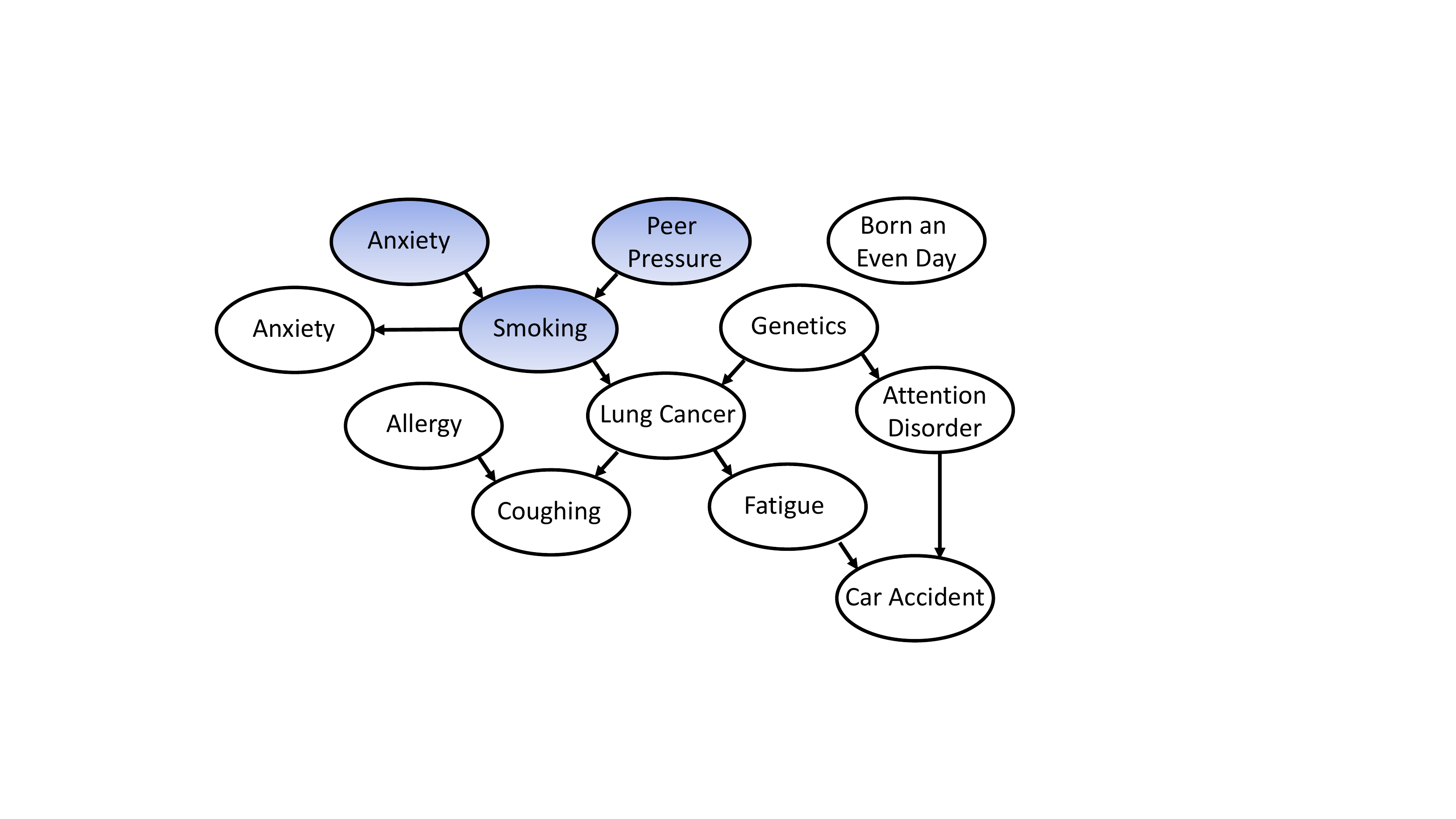}
	\caption{\bf{The causal DAG used to generate CancerData.}}
	\label{fig:exdag}
\end{figure*}

\begin{exa} \label{ex:coll}

	CancerData~\cite{LUCAS} is a simulated dataset generated according to the causal DAG shown in
	Fig.~\ref{fig:exdag}. In this graph, Smoking is a collider in the path between Peer\_Pressure and Anxiety, i.e.,  $\mb P$: Peer\_Pressure $\rightarrow$ Smoking $\leftarrow$ Anxiety. Furthermore, $\mb P$ is the only path between  Peer\_Pressure and Anxiety. Since $\mb P$ is a closed path, Anxiety and Peer\_Pressure
	are marginally independent. This independence holds in CancerData, since $I(\text{Anxiety}, \text{Peer\_Pressure})=0.000004$, which is not statistically significant (pvalue>0.6). Now, since
	Smoking is a collider in $\mb P$, conditioning on Smoking renders spurious correlation between Anxiety and \_Pressure. From CancerData we obtain that, $I(\text{Anxiety}, \linebreak \text{Peer\_Pressure}|  \text{Smoking})=0.003$, which is statistically significant (pvalue<0.001).

\end{exa}

\begin{defn} \label{def:gi} A distribution $\Pr(\mb{A})$ on the
  variables $\mb{A}$ is {\em causal} or {\em DAG-isomorphic} if there
  exists a DAG $G$ with nodes $\mb{A}$ such that\footnote{ The
    $\Rightarrow$ direction is called {\em Causal Markov Assumption}
    and $\Leftarrow$ is called {\em Faithfulness}.}
  $\mb{X} \indep \mb{Y} |_d \ \mb{Z} \Leftrightarrow \mb{X} \indep
  \mb{Y} | \mb{Z}$ \cite{pearl2014probabilistic,
    pearl2009causality,spirtes2000causation}.
\end{defn}

Fix a treatment $T$ and outcome $Y$.  A set $\mb Z$ is said to satisfy
the {\em back-door criterion} if it closes all back-door paths from
$T$ to $Y$.  Pearl \cite{pearl1993bayesian} proved the following:
\begin{theo}~\cite[Th. 3.2.5]{pearl2009causality} \label{theo:par_cov}
  if $\mb Z$ satisfies the back-door criterion, then it satisfies
  Unconfoundedness.
\end{theo}

\paragraph*{\bf Total and Direct Effects} $\ate$ measures the total
effect of $T$ on $Y$, aggregating over all directed paths from $T$ to
$Y$.  In some cases we want to investigate the {\em direct effect}, or
natural direct effect, $\nde$~\cite{pearl2001direct}, which measures
the effect only through a single edge from $T$ to $Y$, which we review
here.  A node $M$ that belongs to some directed path from $T$ to $Y$
is called a {\em mediator}.  We will assume that each unit in the
population has two attributes $Y(t_1)$ and $Y(t_0,M(t_1))$,
representing the outcome $Y$ when we apply the treatment $t_1$, and
the outcome when we don't apply the treatment {\em and simultaneously}
keep the value of all mediators, $\mb{M}$, to what they were when the
treatment $t_1$ was applied.  Then:
\begin{align} \nde(T,Y) \defeq \E[Y(t_0, \mb{M}(t_1))] - \E[Y(t_1)] \label{eq:nde} \end{align}
For example, in gender discrimination the question is whether gender
has any direct effect on income or hiring \cite{pearl2001direct}.
Here $t_1=$Male, $t_0=$Female, $Y$ is the decision to hire, while the
mediators $\mb{M}$ are the qualifications of individuals.  The outcome
$Y(t_0,M(t_1))$ is the hiring decision for a male, if we changed his
gender to female, but kept all the qualifications unchanged.  Since
$Y(t_0,M(t_1))$ is missing in the data, $\nde$ can not be estimated,
even with a controlled experiment \cite{pearl2001direct}.  However,
for mediators $\mb{M} \defeq \mb{PA}_Y -\{T\}$ and covariates
$\mb{Z} \defeq \mb{PA}_T$, it satisfies the {\em mediator formula}
\cite{pearl2001direct}, given by  Eq.(\ref{eq:medi}) in Sec.~\ref{sec:peri}.
%

\subsection{Additional Proofs, Algorithms, Examples and Graphs}

In this section we present some of the proofs and algorithms that were missing in the main part of the paper. We also
present additional examples and graphs.

\paragraph*{\bf Proof of Prop.\ref{def:biasedq}}
%
  We prove (a) (the proof of (b) is similar and omitted).
	Denoting $\mb Y_j \defeq (Y_j(t_0), Y_j(t_1))$, we will prove
	independence in the context $\Gamma_i$, i.e.
	$(\mb Y_j \bigCI T | \Gamma_i)$.  Using (1) unconfoundedness for
	$\mb Z$, $(\mb Y_j \bigCI T | \mb Z=\mb z)$, and (2) the balanced
	assumption for $\mc{Q}$, $(T \indep \mb{Z}| \Gamma_i)$, we show:
	$\E[T | \mb Y_j = \mb y, \Gamma_i] = \E_{\mb z}[\E[T | \mb Y_j = \mb
	y, \Gamma_i, \mb Z = \mb z] | \mb Y_j = \mb y, \Gamma_i] = \E_{\mb
		z}[\E[T | \mb Z = \mb z, \Gamma_i] | \mb Y_j = \mb y, \Gamma_i]$
	(by (1)) $= \E_{\mb z}[\E[T|\Gamma_i] | \mb Y_j = \mb y, \Gamma_i]$
	(by (2)) $= \E[T|\Gamma_i]$ (because $\E[T|\Gamma_i]$ is independent
	of $\mb{z}$), proving $(\mb Y_j \bigCI T | \Gamma_i)$.

\vspace*{0.2cm}

\paragraph*{\bf Completing Example~\ref{ex:simp}}
Listing \ref{rfqex1} shows the rewritten query associated to the
biased query in Ex.~\ref{ex:simp}. Query rewriting removes bias
resulted from the influence of Airport, Year, Day and Month. See
Sec.~\ref{sec:rbias} for details.

\paragraph*{\bf Algorithm for Fine-Grained Explanations} The details of the procedure proposed in Sec.~\ref{sec:ebias} for generating fine-grained explanations for a biased query is shown in  Algorithm \ref{alg:fgealg}.

\begin{figure}
	\begin{subfigure}{0.49\textwidth}
		\includegraphics[height=4cm,width=1\linewidth]{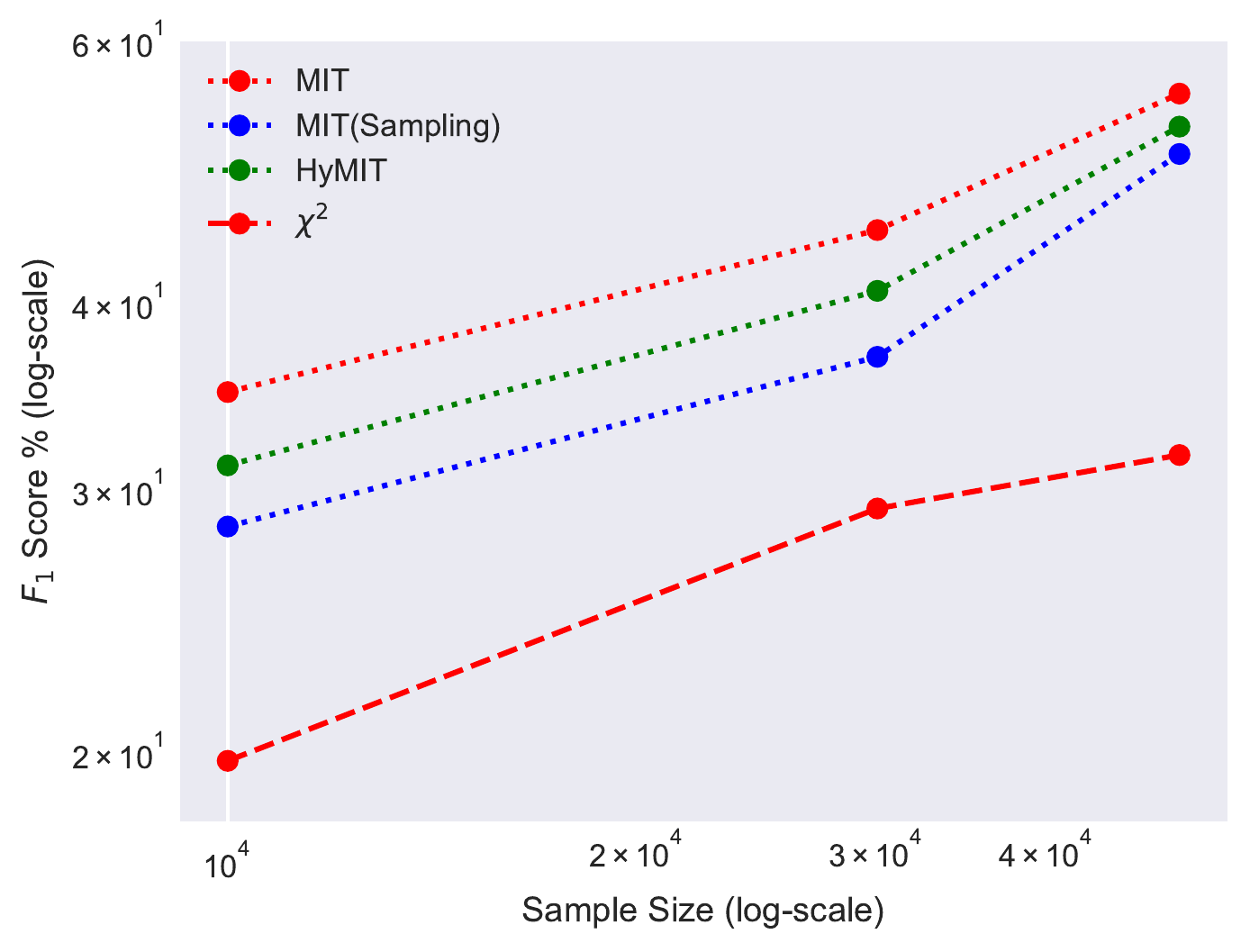}
		\caption{Quality of the optimizations proposed for independence test.}
	\end{subfigure}
\begin{subfigure}{0.49\textwidth}
	\hspace*{0.3cm}	\includegraphics[height=4cm,width=1\linewidth]{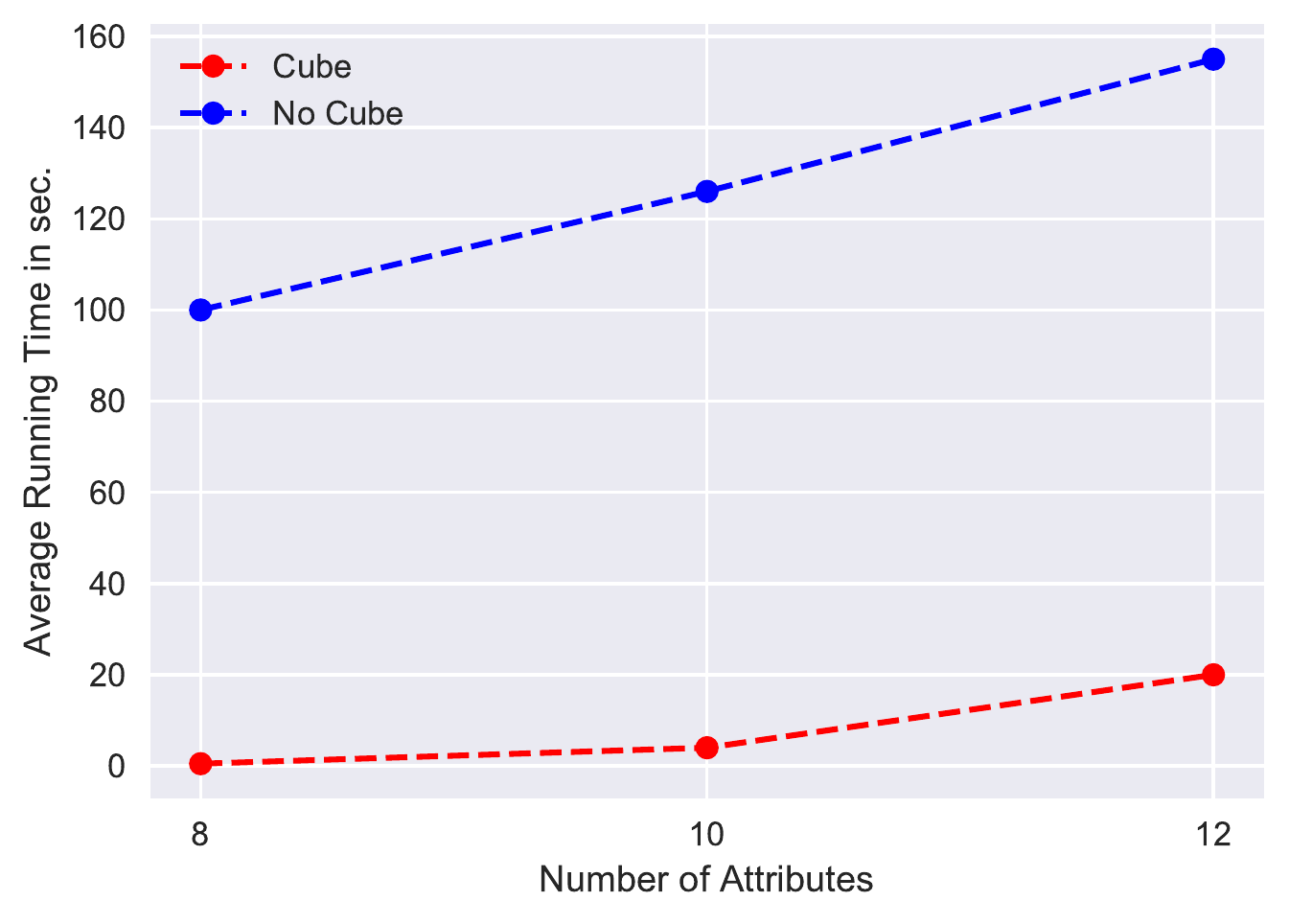}
		\caption{Benefits of using data cubes.}
	\end{subfigure}
	\caption{\bf Additional graphs for experiments in Sec.~\ref{sec:experi}.}
	\label{fig:addplots}
\end{figure}

\paragraph*{\bf Additional Graphs} Figure \ref{fig:addplots} shows additional graphs for the experiments in Sec~\ref{sec:opti_expri}.


\begin{figure}
	\scriptsize
\begin{lstlisting}[language=SQL,escapechar=@,language=SQL, escapeinside={|}{|},
basicstyle=\ttfamily,caption=Rewritten query associated to Ex.~\ref{ex:simp}.,label=rfqex1]
WITH Blocks
AS(
SELECT |Carrier|,|Airport|,|Year|,|Day|,|Month|,avg(|Delay|) AS Avge
FROM FlightData
WHERE |Carrier| in ('AA','UA') AND Airport in ('COS','MFE','MTJ','ROC')
GROUP BY   |Carrier|,|Airport|,|Year|,|Day|,|Month|),
Weights
AS(
SELECT |Airport|,|Year|,|Day|,|Month|, count(*)/n  AS W
FROM FlightData
WHERE Carrier in ('AA','UA') AND Airport in  ('COS','MFE','MTJ','ROC'))
GROUP BY  |Airport|,|Year|,|Day|,|Month|
HAVING count(DISTINCT |Carrier|)=2)
SELECT |Carrier|,sum(Avge * W)
FROM Blocks,Weights
WHERE Blocks.|Airport| = Weights.|Airport| AND
Blocks.|Month| = Weights.|Month| AND
Blocks.|Day| = Weights.|Day| AND
Blocks.|Year| = Weights.|Year| AND
GROUP BY |Carrier|
\end{lstlisting}
\end{figure}
\vspace*{0.5cm}

 \begin{algorithm}
	\DontPrintSemicolon
	\KwIn{ A database $D$, three attributes $T, Y, Z \in \att$, an integer $k$ denotes the number of explanations}
	\KwOut{ Top-$k$ explanations}

	$\mb{S} \gets \emptyset$\;

	\For{$ (t,y,z) \in \Pi_{TYZ}(D)$}
	{
		$K_t[(t,y,z)] \gets 		\kappa_{(t,z)}$\;
		$K_y[(t,y,z)] \gets 		\kappa_{(y,z)}$\;
	}
	$R_i  \gets$ Rank $K_i$ by value, for $i \in \{t,y\}$\;
	$R \gets \mb{RankAggregate}(R_t,R_y)$\;
	\Return  Top-$k$ triples in $R$
	\caption{{\sc Fine-Grained Explanation (\fge)}} 	\label{alg:fgealg}
\end{algorithm}


\end{document}